\documentclass[twocolumn]{aastex7}

\usepackage{tabularx}
\usepackage{longtable}
\usepackage{lineno}
\usepackage{amssymb}
\usepackage{gensymb}
\usepackage{xcolor}

\begin{document}

\linenumbers

\title{Cepheid-based Distances to NGC\,4303 and NGC\,1068}

\author[orcid=0009-0000-9670-2194]{Madison Markham}
\affiliation{Department of Physics and Astronomy, Georgia State University, Atlanta, GA 30303, USA}
\email{mmarkham@gsu.edu}

\author[orcid=0000-0002-2816-5398]{Misty C. Bentz}
\affiliation{Department of Physics and Astronomy, Georgia State University, Atlanta, GA 30303, USA}
\email{bentz@gsu.edu}

\author[orcid=0000-0002-8224-1128]{Laura Ferrarese}
\affiliation{NRC-Herzberg Astronomy and Astrophysics, 5071 West Saanich Road, Victoria, BC, V9E 2E7, Canada}
\email{Laura.Ferrarese@nrc-cnrc.gc.ca}

\author[orcid=0000-0003-0017-349X]{Christopher A. Onken}
\affiliation{Research School of Astronomy and Astrophysics, Australian National University, Canberra, ACT 2611, Australia}
\email{christopher.onken@anu.edu.au}

\author[orcid=0000-0001-9191-9837]{Marianne Vestergaard}
\affiliation{DARK, Niels Bohr Institute, University of Copenhagen, Jagtvej 155, DK-2200 Copenhagen}
\affiliation{Steward Observatory and Department of Astronomy, University of Arizona, 933 N Cherry Avenue, Tucson AZ 85721, USA}
\email{mvester@nbi.ku.dk}

\begin{abstract}

We present Cepheid-based distances to two canonical AGN: NGC\,4303 (M\,61) and NGC\,1068. Data were obtained using the \textit{Hubble Space Telescope} with nonredundant time spacing over 12 visits for each target, and observations were made with the F555W and F814W filters. We found 32,694 point sources in NGC\,4303, and 130 of these were determined to be strong Cepheid candidates with periods ranging $\sim 13-93$ days. In NGC\,1068, we found 20,207 point sources, where 51 of these were strong Cepheid candidates with periods $\sim 14-92$ days. We fit the period$-$luminosity relationship, calibrated based on a geometric distance to the LMC by \cite{riess_large_2019}, to our Cepheid candidates in each galaxy and correct for potential effects of metallicity. Using a distance constraint for the LMC given by \cite{pietrzynski_distance_2019}, this yields a distance modulus of $\mu = 31.083 \pm 0.035$ mag for NGC\,4303 and $\mu = 30.150 \pm 0.106$ mag for NGC\,1068. Thus, we measure distances of $D = 16.47 \pm 0.27$ Mpc to NGC\,4303 and $D = 10.72 \pm 0.52$ Mpc to NGC\,1068.

\end{abstract}


\section{Introduction} \label{sec:intro}

Henrietta Leavitt's work with Cepheid variable stars \citep{leavitt_periods_1912} remains some of the most influential in extragalactic astronomy to date. Her discovery of the period$-$luminosity relationship (also known as the Leavitt Law) allowed for the distance to the Andromeda galaxy to be measured for the first time \citep{hubble_cepheids_1925}, unlocking the distance scale of the Universe. Over a century later, modern extragalactic studies have been revolutionized by the success of the \textit{Hubble Space Telescope} (\textit{HST}). The \textit{HST} Key Project was one significant endeavor with the goal of determining the Hubble constant using Cepheids \citep{freedman_final_2001}. Recently, Cepheids have been observed using \textit{HST} in numerous successful studies to obtain accurate distances to nearby active galactic nuclei (AGN; e.g., \citealt{bentz_cepheid-based_2019, yuan_cepheid_2020, yuan_cepheid_2021}), Type Ia Supernova (SN Ia) host galaxies (e.g., \citealt{saha_cepheid_2006,riess_comprehensive_2022}), and the Magellanic Clouds (e.g., \citealt{riess_large_2019, breuval_small_2024}). 

In addition to its importance in Cepheid studies, one of the more surprising discoveries by \textit{HST} was that all massive galaxies likely host supermassive black holes (SMBHs) at their centers. This has inspired a wealth of research on SMBHs and galactic evolution, and SMBHs have been found to coevolve with their hosts (e.g., \citealt{kormendy_coevolution_2013}). Correlations of SMBH mass with bulge velocity dispersion ($M_{\rm BH} - \sigma_{\star}$; \citealt{ferrarese_fundamental_2000, gebhardt_relationship_2000, gultekin_m-ensuremathsigma_2009}) and bulge luminosity ($M_{\rm BH} - L_{\rm bulge}$; \citealt{kormendy_inward_1995, kormendy_coevolution_2013}) are interpreted to mean that active black holes affect their host galaxies on massive scales, far greater than their gravitational influence. AGN feeding and feedback processes seem to play an essential role in galactic evolution, and SMBHs lie at the centers of these systems. 

Accurate distances to nearby AGN are some of the most crucial measurements to be made of these objects, and large uncertainties in distance measurements inhibit useful measurements of several other important characteristics, such as central black hole mass. Many direct black hole mass measurements (which have been made for both targets in our sample) rely on spatial resolution, and this is directly related to distance. Unlike AGN at cosmological distances, nearby AGN are not all reliably within the Hubble Flow; thus, redshift-independent distances are required. NGC\,4151, for example, is one of the most widely studied Seyfert 1 AGN; it is an incredibly valuable observational resource for feeding and feedback studies as one of the nearest and brightest AGN (e.g., \citealt{ulrich_active_2000}). Its distance was previously quoted as a group average, with four members ranging $3.9 - 34$ Mpc \citep{tully_extragalactic_2009}. Recent work by \cite{yuan_cepheid_2020} was able to constrain the distance to NGC\,4151 to $15.8 \pm 0.4$ Mpc using Cepheids observed with \textit{HST}. 

Seyfert 2 AGN are distinct from Seyfert 1 AGN by a viewing angle-imposed obscuration of the SMBH with a dusty torus. NGC\,1068, in particular, was instrumental in creating a unified AGN model through the discovery of broad Balmer lines in the polarized light from this Seyfert 2 galaxy \citep{antonucci_spectropolarimetry_1985}. High angular-resolution studies of the obscuring molecular gas result in a better understanding of the structure of the torus (e.g. in NGC\,1068, \citealt{gravity_collaboration_image_2020, gamez_rosas_thermal_2022, gamez_rosas_decoding_2025}), and determining physical size from angular size depends directly on distance. Understanding the structure of the torus then allows for better-fitting models for the spectral energy distributions (SEDs) of AGN. Furthermore, describing AGN luminosity with these models requires accurate distances, which NGC\,1068 lacks. Another canonical Seyfert 2, NGC\,4303, is a popular target for spatial studies on star formation in spiral galaxies\,---\,in particular, it is included in the Physics at High Angular resolution in Nearby GalaxieS (PHANGS) program \citep{schinnerer_physics_2019}. \cite{emsellem_phangs-muse_2022} describe the need for understanding characteristics such as star formation timescales and stellar feedback processes at spatially-resolved scales in nearby galaxies, and the PHANGS-MUSE survey is working to address this. However, the assumed distance for NGC\,4303 directly affects the spatial scale of star formation in this galaxy, among other things. We therefore set out to rectify the lack of accurate distances to these two canonical AGN by using \textit{HST} observations to detect and characterize Cepheids in their host galaxies.

\begin{figure*}[htb!]
    \centering
    \includegraphics[width=\columnwidth]{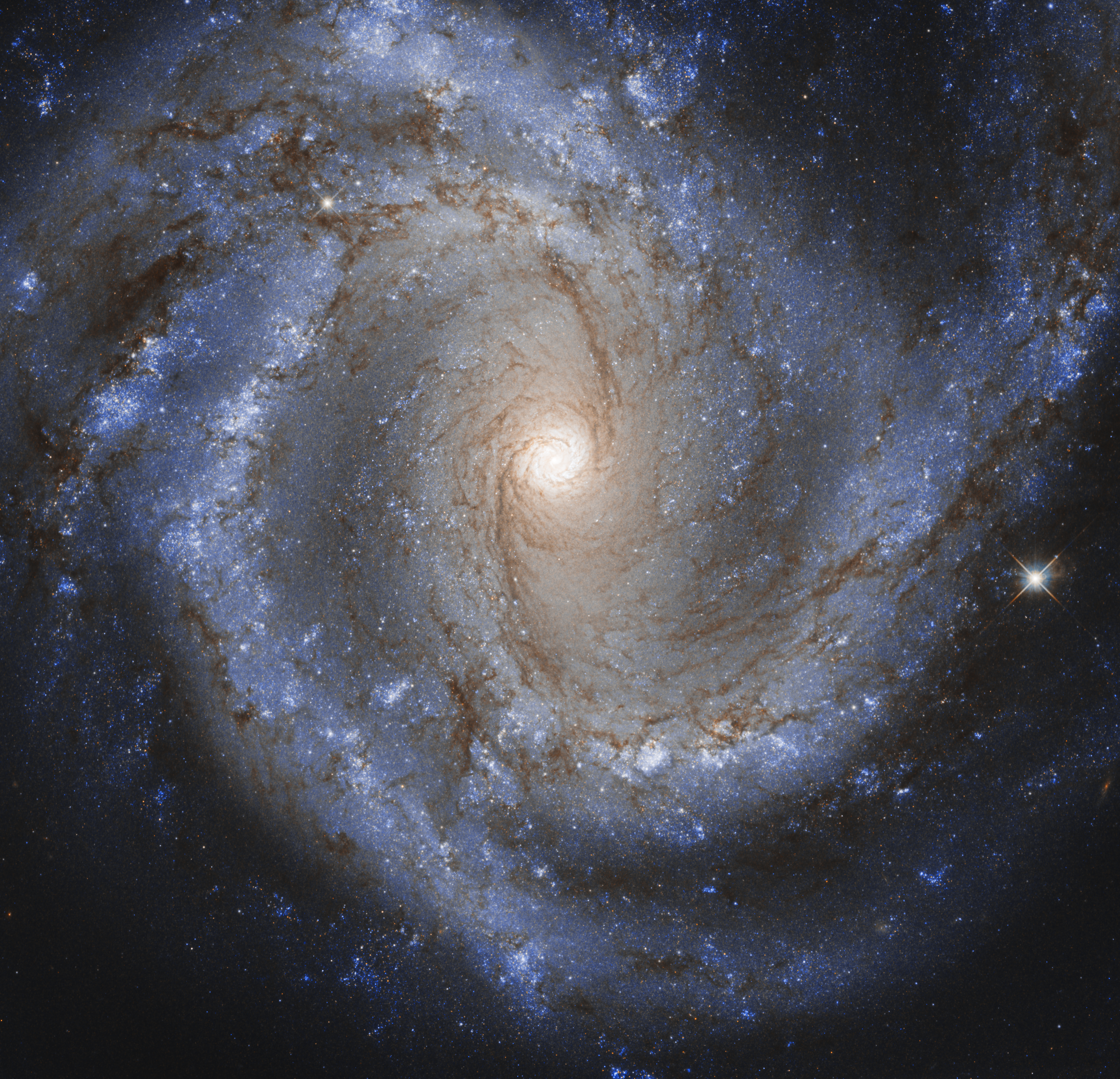}
    \includegraphics[width=0.832\columnwidth]{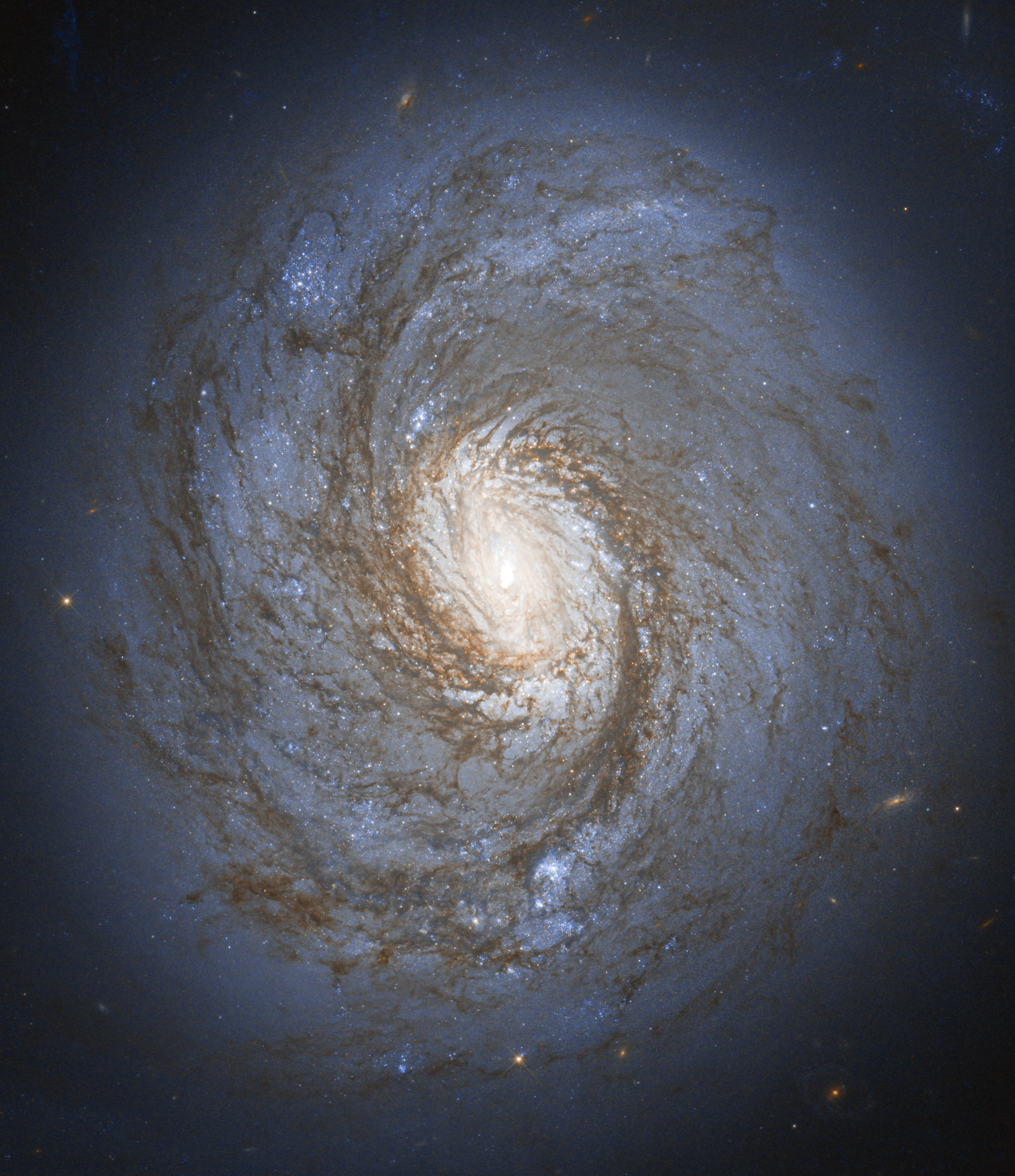}
    \caption{Red (F814W), blue (F555W), and pseudogreen color images of NGC\,4303 (left) and NGC\,1068 (right) using \textit{HST} Wide Field Camera 3 data. The field of view of NGC\,4303 here is $2.6' \times 2.6'$. This image of NGC\,1068 has been cropped to remove a foreground star, and its field of view is $2.1' \times 2.6'$. NGC\,4303 is oriented with North approximately up, and NGC\,1068 has North oriented $29.22 \degree$ clockwise from up. Credit: Judy Schmidt.}
    \label{fig:gals}
\end{figure*}

NGC\,4303 (also known as M\,61) is a late-type Sbc galaxy with a double bar and a nuclear starburst region (\citealt{binggeli_studies_1985, schinnerer_toward_2002}; see Figure \ref{fig:gals}). It is classified as a Seyfert 2 AGN, and \cite{pastorini_supermassive_2007} derive a central black hole mass of $M_{\rm BH} = 5.0^{+0.9}_{-2.3} \times 10^6 \ M_{\odot}$ using STIS observations of ionized gas kinematics. These authors assume a distance of 16.1 Mpc (\citealt{schinnerer_toward_2002}, using the results of \citealt{ferrarese_extragalactic_1996}), which corresponds to a scale of $1'' \approx 78$ pc. Redshift-independent distance measurements have primarily been obtained using Tully$-$Fisher and Type II supernovae (SNe II) methods, and these measurements span a range of $\sim7-20$ Mpc \citep{bottinelli_hi_1984, rodriguez_photospheric_2014}. NGC\,4303 is nearly face-on ($i=25\degree$ from $^{12}$CO velocity fields;  \citealt{schinnerer_toward_2002}), which contributes to large uncertainties in distances derived using the Tully$-$Fisher method. Additionally, though it has hosted an incredible number of supernovae (8 known since 1926!), none of them have been Type Ia and suitable as standard candles. However, two of those supernovae were Type IIP supernovae (SNe IIP), and there have been recent attempts to use SNe IIP as standardizable candles \citep{csornyei_consistency_2023}. \cite{anand_distances_2021} attempted to constrain its distance using the tip of the red giant branch (TRGB) method, however their results were inconclusive. NGC\,4303 has been studied in the context of the starburst$-$AGN connection (e.g., \citealt{riffel_sinfoni_2016, dametto_sinfoni_2019}), though the results are limited by the large uncertainties in distance. Furthermore, the association of NGC\,4303 with the Virgo cluster is uncertain \citep{binggeli_studies_1985}, and therefore the influence of the Virgo cluster on the star formation rate in NGC\,4303 is also uncertain. An accurate distance is necessary to better understand feeding and feedback processes between the AGN and nuclear starburst region.

NGC\,1068 is one of the best-studied Seyfert galaxies; it belongs to the original sample of 12 Seyfert galaxies \citep{seyfert_nuclear_1943}, and it is often described as the prototypical Seyfert 2 AGN. NGC\,1068 is an Sb galaxy with a weak bar visible in the NIR and a circumnuclear starburst ring (\citealt{telesco_luminous_1984,scoville_stellar_1988,thronson_near-infrared_1989,davies_star_1998}; see Figure \ref{fig:gals}). \cite{lodato_non-keplerian_2003} constrained its black hole mass to $M_{\rm BH} = 8.0 \pm 0.3 \times 10^6 \ M_{\odot}$ using water maser kinematics, assuming a distance of 14.4 Mpc and a scale of $1'' \approx 69$ pc. The disk is moderately inclined, having an inclination angle of $i=40 \pm 3 \degree$ from observations of \ion{H}{1} gas kinematics \citep{brinks_hi_1997}. Famously, NGC\,1068 is also an archetype of AGN unification schemes (e.g., \citealt{antonucci_spectropolarimetry_1985, konigl_disk-driven_1994}). NGC\,1068 is therefore a fantastic laboratory for studies of feeding and feedback processes, specifically through emission-line observations in the central disk. AGN-driven outflows of dense molecular gas have been revealed (e.g., \citealt{garcia-burillo_molecular_2014}), and this has inspired studies on the structure and kinematics of molecular gas in the torus of NGC\,1068, such as recent work by \cite{gamez_rosas_decoding_2025}. Despite this, its distance has been neglected. Similar to NGC\,4303, its distance has primarily been constrained using the Tully$-$Fisher method, with many studies adopting 14.4 Mpc from \cite{tully_catalog_1988}. Recently, \cite{tikhonov_trgb_2021} determined a TRGB distance of $11.14 \pm 0.54$ Mpc, though this was accomplished using inhomogeneous archival \textit{HST} data. A more accurate distance to NGC\,1068 is thus needed for crucial measurements of AGN feedback processes.

This paper is organized as follows. Section \ref{sec:obs} provides descriptions of the observations for each target. Section \ref{sec:phot} outlines the photometric techniques employed to determine light curves for all point sources, while Section \ref{sec:cand} explains our criteria for selecting Cepheid candidates among these sources. In Section \ref{sec:pl}, we fit the resulting period$-$luminosity relationships, and we considered the potential effects of metallicity. We discuss previous distance measurements to each target in Section \ref{sec:disc}, as well as the implications of our Cepheid-based distances. Finally, Section \ref{sec:sum} provides a summary.

\section{Observations} \label{sec:obs}

Multi-epoch observations for each target were obtained with the \textit{HST} Wide Field Camera 3 (WFC3)\footnote{The specific observations analyzed can be accessed from MAST at the Space Telescope Science Institute via \dataset[doi: 10.17909/ht8k-kg64]{https://doi.org/XYZ10.17909/ht8k-kg64}.}. We observed in two optical bandpasses (F555W \& F814W) with the UVIS channel, which has a field of view of $2.6' \times 2.6'$ and a pixel scale of $0.04''$. We selected the UVIS-FIX aperture to ensure a fixed field center and aid in data analysis. Similarly, we planned the visits with a fixed roll angle. Observations in each filter employed a four-point dither pattern to increase the sampling of the point spread function (PSF) and fill in the physical gap between the two UVIS detectors.

Cepheids are known to have greater pulsation amplitudes at bluer wavelengths; however, bluer wavelengths are more susceptible to extinction \citep{freedman_hubble_1994}. Thus, the F555W filter maintained a good compromise for maximizing variability while minimizing reddening effects. We also gathered observations in the F814W filter, which is less sensitive to variability and minimally affected by reddening. This provided us with color information for selecting Cepheid candidates.

Observations of NGC\,4303 were taken between April and June of 2023, spanning a total of 70.93 days. NGC\,1068 was observed between December 2023 and February 2024 over a period of 70.61 days. Further details of our observing cadence can be found in Table \ref{tab:obslog}.

\begin{deluxetable*}{cccc}
    \tablewidth{0pt}
\tablecaption{Observations}
\tablehead{
\colhead{Visit} &
\colhead{UT} &
\colhead{Filter} &
\colhead{Exposure Time (s)}
}

\startdata
\multicolumn{4}{c}{NGC\,4303} \\
\hline
$1$ & 2023 Apr 19 & F814W & 2356 \\
$2$ & 2023 Apr 30 & F555W & 2364 \\
$3$ & 2023 May 6 & F555W & 2364 \\
 & \nodata & F814W & 2356 \\
$4$ & 2023 May 11 & F555W & 2364 \\
$5$ & 2023 May 14 & F555W & 2364 \\
$6$ & 2023 May 17 & F555W & 2364 \\
$7$ & 2023 May 19 & F555W & 2364 \\
 & \nodata & F814W & 2356 \\
$8$ & 2023 May 21 & F555W & 2364 \\
$9$ & 2023 Jun 3 & F555W & 2364 \\
 & \nodata & F814W & 2356 \\
$10$ & 2023 Jun 10 & F555W & 2364 \\
$11$ & 2023 Jun 20 & F555W & 2364 \\
 & \nodata & F814W & 2356 \\
$12$ & 2023 Jun 29 & F555W & 2364 \\
\hline
\multicolumn{4}{c}{NGC\,1068} \\
\hline
$1$ & 2023 Dec 19 & F555W & 2364 \\
 & \nodata & F814W & 2356 \\
$2$ & 2024 Jan 5 & F814W & 2356 \\
$3$ & 2024 Jan 10 & F555W & 2364 \\
$4$ & 2024 Jan 14 & F555W & 2364 \\
$5$ & 2024 Jan 19 & F555W & 2364 \\
 & \nodata & F814W & 2356 \\
$6$ & 2024 Jan 21 & F555W & 2364 \\
$7$ & 2024 Feb 2 & F555W & 2364 \\
 & \nodata & F814W & 2356 \\
$8$ & 2024 Feb 10 & F555W & 2364 \\
$9$ & 2024 Feb 17 & F555W & 2364 \\
 & \nodata & F814W & 2356 \\
$10$ & 2024 Feb 23 & F555W & 2364 \\
$11$ & 2024 Feb 25 & F555W & 2364 \\
$12$ & 2024 Feb 28 & F555W & 2364 \\
\label{tab:obslog}
\enddata
\end{deluxetable*}

We scheduled 12 total visits with nonredundant time spacing. For each target, images were taken in the F555W filter, and five of these visits included F814W observations adopting the same position and orientation. In the case of NGC\,4303, two of the visits did not achieve guide star lock, and one of these two visits was re-observed at the end of the monitoring sequence. The visit that was re-observed required us to adopt a different roll angle, and we chose to minimize the difference between the new roll angle and our previous roll angle to ensure the maximum possible overlap of the FOV for the final visit compared to the previous visits. Observations of NGC\,1068 were plagued by similar issues. Four visits did not achieve guide star lock for the F555W observations, and three of these were rescheduled for the end of the monitoring sequence. These visits also required us to adopt different roll angles, and these angles were also kept as close to the original roll angle as possible.  

To prepare our observations for analysis of the photometry, we downloaded charge transfer efficiency-corrected images from the Mikulski Archive for Space Telescopes (MAST). The images for each target were registered and drizzled to a common frame using the AstroDrizzle package \citep{fruchter_betadrizzle_2010}. First, all observations in F555W were stacked to create a single ``deep frame" image for each galaxy. This ``deep frame" was then used as a reference for drizzling each of the individual visits in both filters separately. This ensured that all of our images were aligned to a common coordinate frame, so point sources could easily be matched between individual visits.

\subsection{Photometry} \label{sec:phot}

In general, we followed the same photometric procedure that is outlined in \cite{bentz_cepheid-based_2019}. For each target, we identified point sources in the field and performed point-spread function (PSF) photometry using DAOPHOT \citep{stetson_daophot_1987}.

Beginning with NGC\,4303, we used the drizzled F555W deep frame to identify all point sources. We then identified a sample of 89 relatively bright and spatially distinct sources to create a PSF model. These sources were carefully examined for consistency in the shapes of their individual PSFs, so broad profiles, extended sources, and sources with close neighbors were excluded. The resulting PSF model was used to perform PSF photometry on all point sources in the F555W deep frame. Next, the same sample of 89 stars was used to create PSF models for all of the individual visits in the F555W filter. We selected a new sample of 82 sources to create a PSF model in the F814W filter and all corresponding individual visits. For NGC\,1068, we repeated this procedure. We used a sample of 64 stars to construct our PSF model in F555W, and we used a sample of 61 stars in F814W. Aperture corrections for the PSF photometry were derived by calculating the differences between the fitted PSF magnitude and a standard WFC3 10-pixel aperture magnitude for all sources used to create the PSF model. The median difference was accepted as the aperture correction, which yielded brighter corrected magnitudes. Thus, we had unique aperture corrections in F555W and F814W for each target, which were typically found to be around $-0.2$ mag and $-0.3$ mag, respectively.

For each galaxy, point sources were matched between visits in F555W and F814W. This yielded 32,694 unique point sources in our field for NGC\,4303. Matching point sources similarly in NGC\,1068 yielded 20,207 unique point sources. One potential explanation for the increased number of sources in NGC\,4303 is the greater star formation rate compared to NGC\,1068. Finally, we adopted the magnitude zeropoints for a WFC3 10-pixel aperture given by \cite{riess_large_2019}: 25.727 mag for F555W and 24.581 mag for F814W in the Vegamag system. In NGC\,4303, the typical $10\sigma$ depth given the median uncertainty for an individual visit was $\sim 26.3$ mag in F555W and $\sim 25.2$ mag in F814W. In NGC\,1068, these typical $10\sigma$ depths were $\sim 26.5$ mag in F555W and $\sim 25.3$ mag in F814W.

\section{Cepheid Candidate Selection} \label{sec:cand}

\begin{figure}[htb!]
    \centering
    \includegraphics[width=\columnwidth]{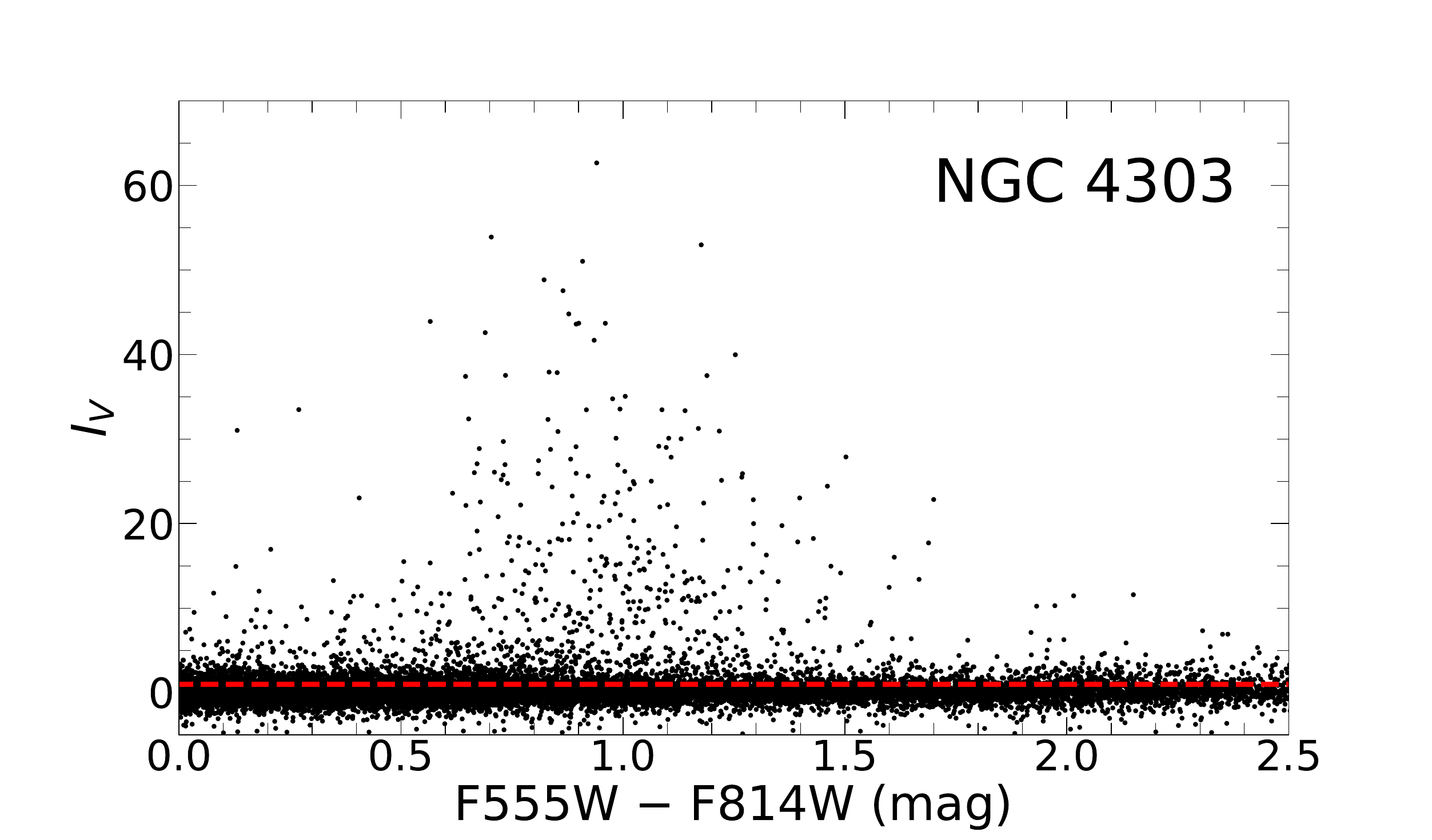}
    \includegraphics[width=\columnwidth]{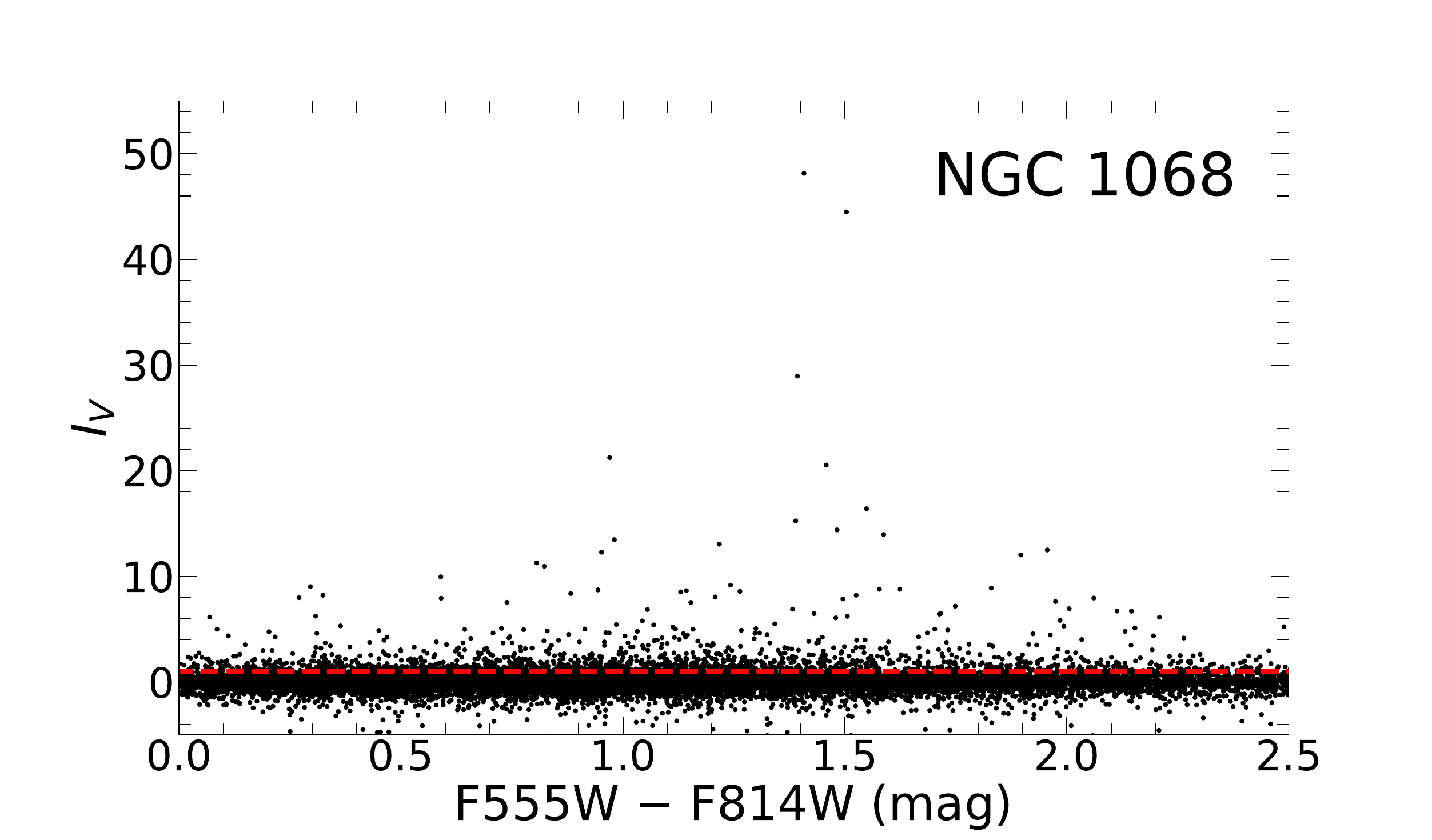}
    \caption{Variability index $I_V$ for point sources in NGC\,4303 (top) and NGC\,1068 (bottom), shown against color. The red dashed line represents $I_V = 1.0$, so all sources above these lines were considered to be variable candidates.}
    \label{fig:Ivar}
\end{figure}

To begin our search for variable candidates, we calculated the \cite{welch_robust_1993} variability index for each point source, given by:
\begin{equation}
    I_{V} = \sqrt{\frac{1}{n(n-1)}} \sum^{n}_{i=1} \delta V_{i} \delta I_{i}
    \label{eq:var}
\end{equation} where $\delta V_{i}$ and $\delta I_{i}$ are the magnitude deviations in F555W and F814W, respectively, relative to the weighted mean magnitudes. The strength of this method comes from the fact that photometric errors should be uncorrelated between images in separate bandpasses. Our criteria for variable objects was a variability index of $I_{V} > 1.0$, displayed in Figure \ref{fig:Ivar}. We chose this criteria to be generous (compared to \cite{shappee_new_2011}, for example), as later cuts to other parameters would remove additional unlikely candidates. Based on this selection, we were left with 4,540 variable candidates in NGC\,4303 and 1,885 variable candidates in NGC\,1068.

Once the variable sources had been identified, we fit the long-period Cepheid templates by \cite{yoachim_panoply_2009} to the respective light curves. Following the procedure of \cite{bentz_cepheid-based_2019}, we injected the fitting routine with a series of 150 initial periods spaced logarithmically between 10 and 100 days. Sources that are true Cepheid candidates should converge to a best-fit period for closely spaced initial guesses near the true period. Each variable source was fit with LMC long period ($>$10 days) Cepheid templates. Strict limits were set so that periods $<$10 days or $>$100 days were not allowed. This ensured that we were not accidentally fitting short-period Cepheids (since we are selecting for long-period Cepheids) and that our observations spanned at least $70\%$ of the period. The successful light curve fits were ranked by $\chi^2$ value, and we examined the six fits with the lowest $\chi^2$ values for each target.  We excluded targets where the six best fits found periods that differed by more than 0.3 days. We also excluded those with $\chi^2$ values of straight line fits that were lower than the $\chi^2$ values for the best LMC long-period Cepheid template fit. 

In NGC\,4303, there were 579 total sources that passed these cuts, spanning a best-fit period range of $10.00 \leq P \leq 93.27$ days. NGC\,1068 had 246 sources pass, with a best-fit period range of $10.00 \leq P \leq 93.18$ days. For each source, we recorded the period, phase, mean F555W magnitude, and mean F814W magnitude returned by the best-fit light curve parameters. 

We then analyzed the colors and amplitudes of these selected targets. Color selections were made based on the expected color of Cepheids on the instability strip \citep{shappee_new_2011}. We adopted the range $0.5 < \rm F555W - F814W < 2.0$, chosen again to be generous given the small Galactic extinction along our line of sight to these sources \citep{bentz_cepheid-based_2019}. This reduced the number of sources in NGC\,4303 from 579 to 435, and in NGC\,1068 from 246 to 181. Sources in NGC\,1068 were redder on average than sources in NGC\,4303, which may be influenced by the greater inclination of NGC\,1068. F555W$-$F814W color uncertainties were restricted to be $<$0.3 mag \citep{bentz_cepheid-based_2019}, though this only removed a single source from NGC\,1068. We also examined the F814W/F555W amplitude ratios based on typical values for fundamental mode pulsators \citep{macri_new_2006,shappee_new_2011}. We found that amplitude ratios of F814W to F555W were all between 0.25 and 0.75 mag, so this criterion also did not remove any sources from either galaxy. We visually examined the light curves of the selected candidates for strong Cepheid signatures and excluded candidates that do not exhibit the characteristic ``sawtooth" variability profile. We were then left with 149 Cepheid candidates in NGC\,4303 and 52 Cepheid candidates in NGC\,1068.

\begin{figure*}[htb!]
    \centering
    \includegraphics[width=1.1\columnwidth]{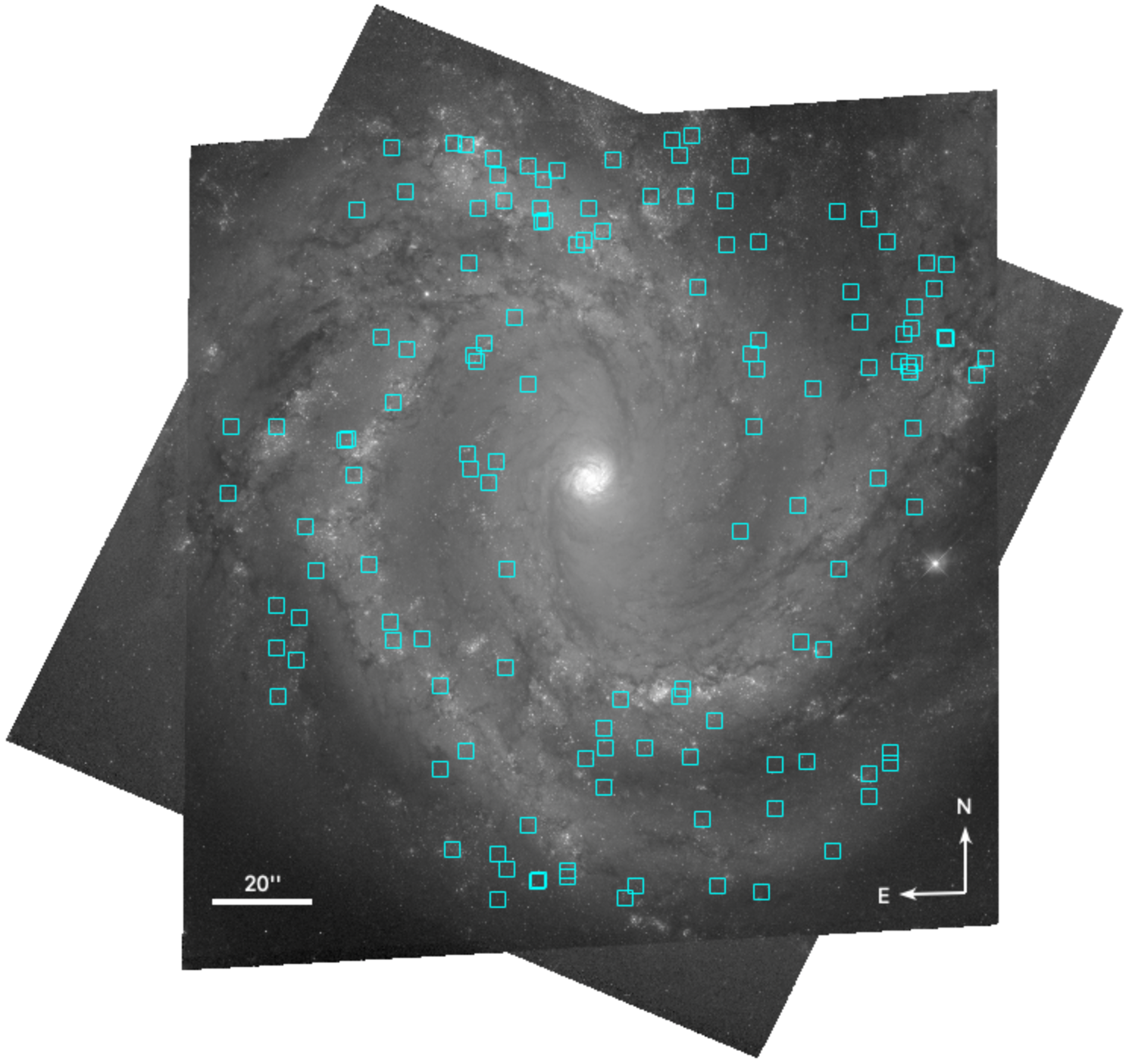}
    \includegraphics[width=0.9\columnwidth]{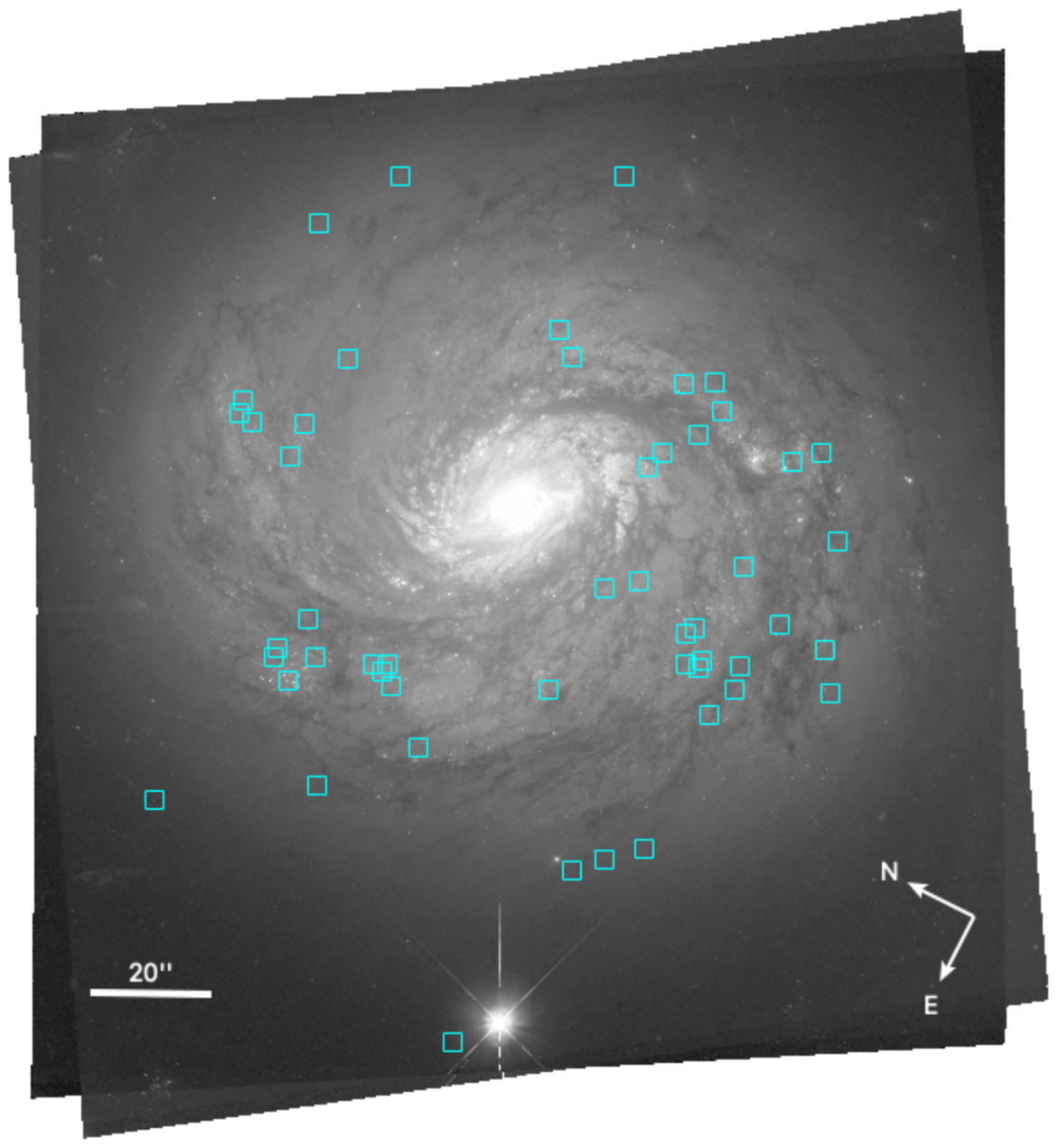}
    \caption{Drizzled ``deep frame" F555W-band images of NGC\,4303 (left) and NGC\,1068 (right) with Cepheid candidate positions marked by cyan boxes. The Cepheid candidates marked here represent our final samples for each galaxy. Compasses are provided in the lower right corners, and $20''$ scale bars are provided in the lower left corners.}
    \label{fig:pos}
\end{figure*}

Finally, we corrected the magnitudes of our remaining Cepheid candidates for the effects of field crowding. Using ADDSTAR, we injected artificial stars randomly into a $40 \times 40$ pixel box surrounding each Cepheid candidate. A single star was injected 100 times, yielding 100 different ``postage stamp" images for each Cepheid candidate. These artificial stars were randomly assigned magnitudes in the range $\pm 0.5$ mag with respect to the Cepheid candidate magnitude. The same general photometric steps were followed as described above to measure the magnitude of all sources, including the artificial star, in each ``postage stamp." The median difference between the injected magnitudes of the artificial stars and the measured magnitudes of the artificial stars was calculated for each Cepheid candidate, and these differences were then averaged to yield the correction required in each filter to remove the effects of crowding. For NGC\,4303, in F555W, the average difference ($m_{\rm injected} - m_{\rm measured}$) was $0.011 \pm 0.044$ mag. In F814W, the average difference was $0.019 \pm 0.043$ mag. For NGC\,1068, the average differences were $0.020 \pm 0.046$ mag and $0.008 \pm 0.033$ mag in F555W and F814W, respectively. Since the average differences between the injected magnitudes and measured magnitudes were positive, crowding effects biased the measured magnitudes slightly brighter. To correct for this, the average differences were added to the mean magnitudes for each Cepheid candidate in the two galaxies.

\section{Period$-$Luminosity Relationship} \label{sec:pl}

For each galaxy, we constructed period$-$luminosity relationships using the final sample of Cepheid candidates in three different bands\,---\,F555W, F814W, and the extinction-free Wesenheit magnitude ($W_I$, \citealt{madore_period-luminosity_1982}). The Wesenheit magnitudes are given by:
\begin{equation}
    W_I = m_{\rm F814W} - R(m_{\rm F555W} - m_{\rm F814W})
    \label{eq:wes}
\end{equation} where $R = A_I / (A_V - A_I)$. We adopted $R=1.3$ from \cite{cardelli_relationship_1989}. Since Wesenheit magnitudes are inherently less susceptible to reddening, there was no need to correct them for Galactic extinction along the line of sight \citep{madore_period-luminosity_1982}.

\cite{riess_large_2019} produced calibrated period$-$luminosity relationships using Cepheids in the LMC for F555W, F814W, and $ W_I$ magnitudes (also using $R=1.3$). Adopting these calibrations eliminated any additional uncertainties that may be introduced by converting \textit{HST} filter magnitudes to \textit{V}- and \textit{I}-band magnitudes. The period$-$luminosity relationship is as follows:  
\begin{equation}
    m = \alpha \, \rm{log} \, \textit{P}\, + \beta
    \label{eq:pl}
\end{equation} where $m$ is the absolute magnitude in each band, $\alpha$ represents the fixed slope from \cite{riess_large_2019} calibrations, $\beta$ represents the intercept in magnitudes, and $P$ is the pulsation period in days.

\begin{figure*}[htb!]
    \centering
    \includegraphics[width=\columnwidth]{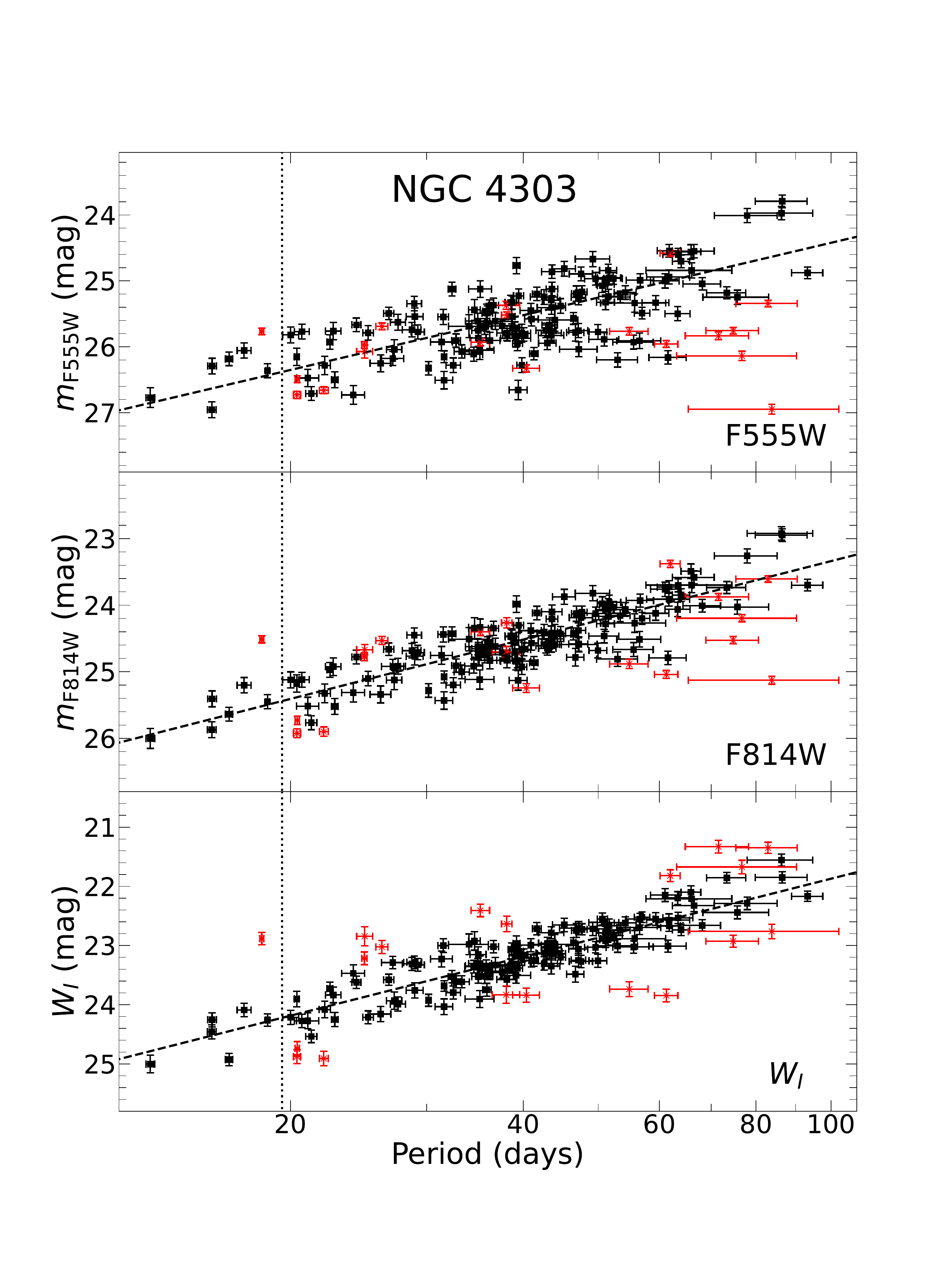}
    \includegraphics[width=\columnwidth]{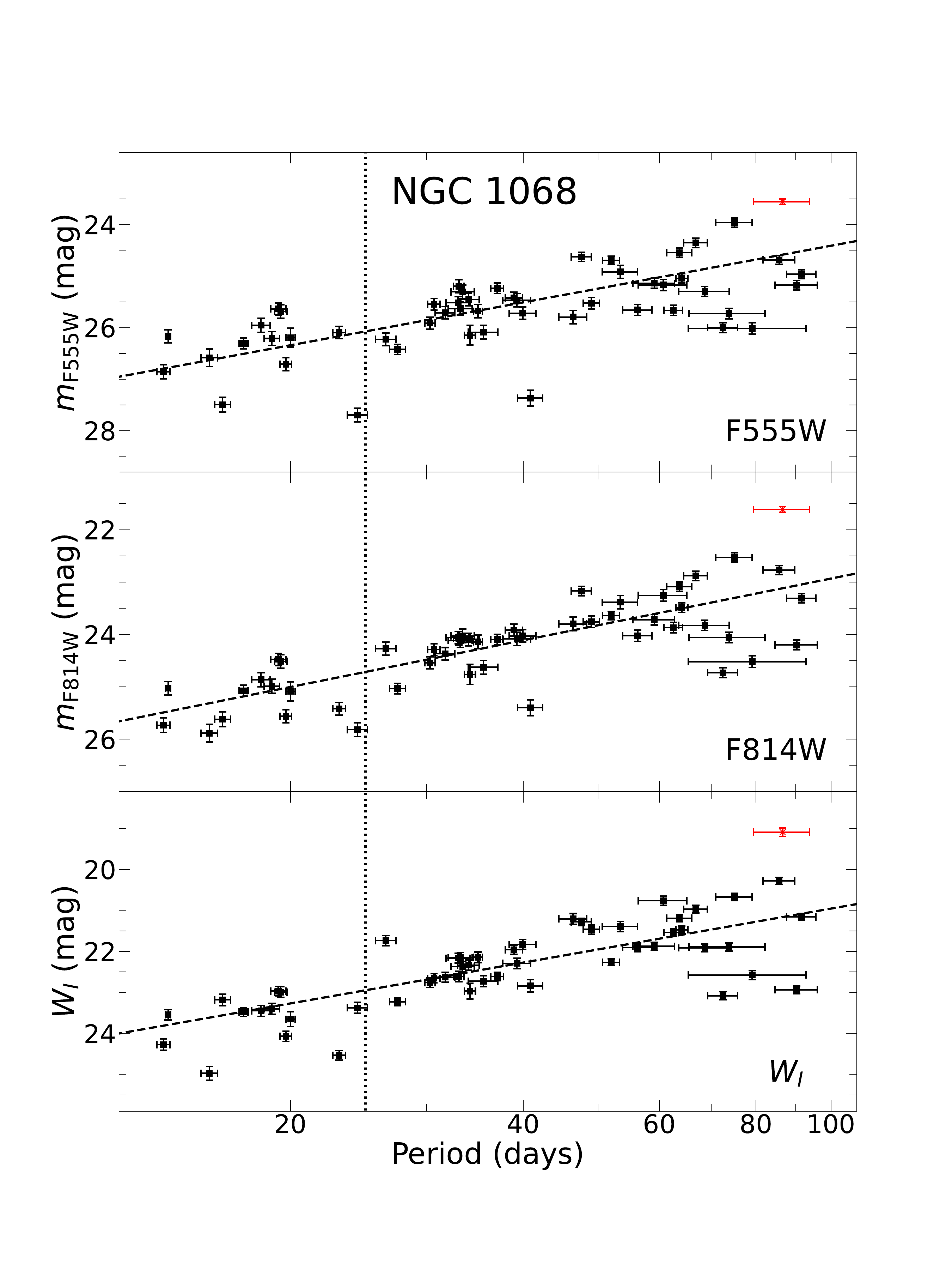}
    \caption{Period$-$luminosity relationships for Cepheid candidates in NGC\,4303 (left) and NGC\,1068 (right). The three panels display F555W (top), F814W (middle), and Wesenheit magnitudes (bottom). Red x's represent Cepheid candidates that were determined to be blue blends or $>3\sigma$ outliers in the Wesenheit index. The dashed lines in each panel represent the lines of best fit, with fixed slopes based on the findings of \cite{riess_large_2019}. The dotted vertical lines represent points of completeness in our sample\,---\,NGC\,4303 was visually determined at $P=19.5$ days, and NGC\,1068 was determined from Figure \ref{fig:comp} at $P=25$ days. Corrections for possible metallicity effects and Galactic extinction have not been applied here.}
    \label{fig:PL}
\end{figure*}

To fit our sample of Cepheids, we used the SciPy ODR routine \citep{virtanen_scipy_2020}. This Levenberg-Marquardt-type algorithm uses an orthogonal distance regression while accounting for uncertainties in both x- and y-dimensions to return best-fit parameters for the fitting function\,---\,in this case, the fitting function was given by Equation \ref{eq:pl}. Holding the slope fixed to the findings of \cite{riess_large_2019}, we found the intercept in each band. Using the residual variance for each fit returned by the SciPy routine, we calculated $\sigma$ in each band. Period$-$luminosity relationships for each band are displayed in Figure \ref{fig:PL}.

We investigated the possibility of blue blends among our sample, as indicated by sources with observed magnitudes significantly brighter than their corresponding predicted magnitudes. Cepheids that lie in the instability strip span a usual $V - I$ color range of $\sim 1$ mag \citep{mochejska_direct_2000}. Cepheid candidates observed to be $> 1$ mag brighter than predicted are likely Cepheids blended with bluer, brighter stars \citep{shappee_new_2011}. Using the same criteria as \cite{bentz_cepheid-based_2019}, we found no Cepheids in NGC\,4303 with $m_{\rm pred} - m_{\rm obs} \geq 1.0$ mag in the F555W filter. We found one likely blue blend candidate in NGC\,1068, and this reduced our sample from 52 to 51 Cepheid candidates. 

\begin{figure*}[htb!]
    \centering
    \includegraphics[width=\columnwidth]{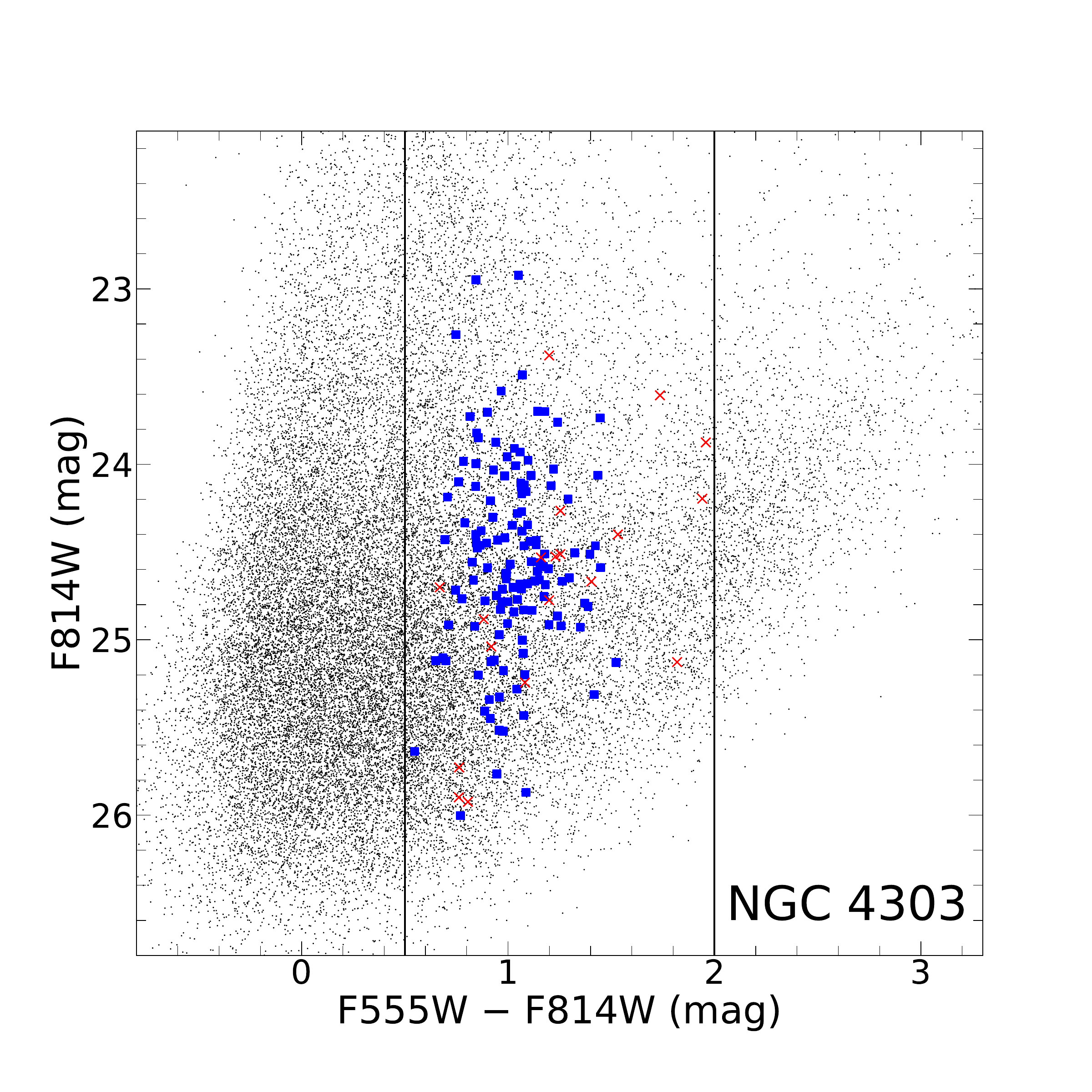}
    \includegraphics[width=\columnwidth]{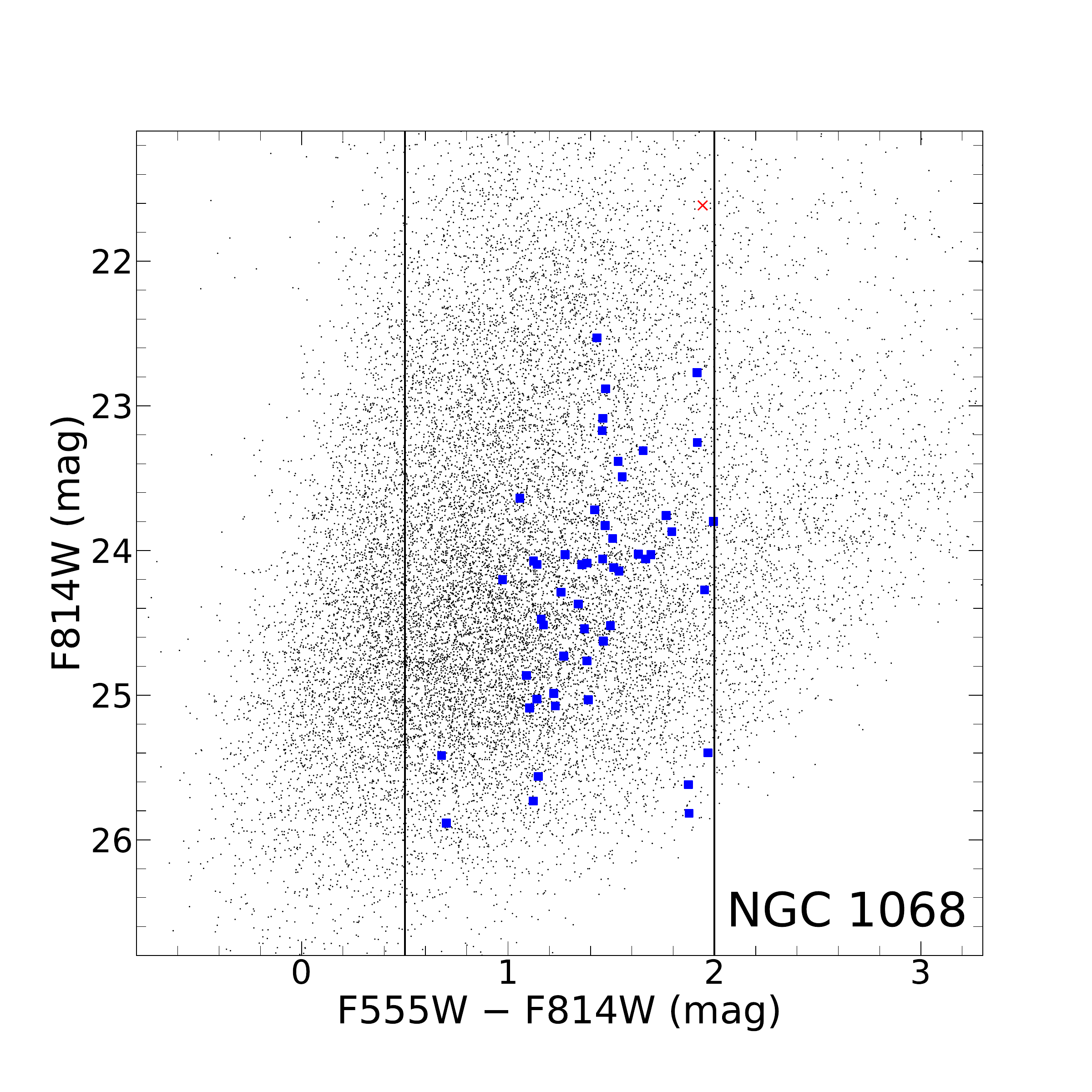}
    \caption{Color$-$magnitude diagrams of all point sources detected in NGC\,4303 (left, shown in black) and NGC\,1068 (right, shown in black). The final samples of Cepheid candidates are marked by blue squares, while the blue blends and $>3\sigma$ outliers are marked by red x's. Solid vertical lines delimit our range of acceptable F555W$-$F814W colors.}
    \label{fig:cm}
\end{figure*}

Outliers in period$-$luminosity relationships can most often be attributed to misidentification\,---\,either of the mode of oscillations or of the class of variable star \citep{kodric_m31_2015}. To account for these outliers, we followed the recommendation of \cite{kodric_m31_2015} and instituted a $3\sigma$-clipping routine that only applied to the $W_I$ magnitudes. Each time the regression was run, the greatest outlier outside of a $3\sigma$ distance was removed and a new fit was determined. This process was done iteratively until there were no more points with $>3\sigma$ distance from the best fit. The outliers were removed from the other filters, and then the final fits in all three bands were determined and recorded. The final sample for NGC\,4303 was reduced from 149 to 130 Cepheid candidates, and for NGC\,1068 the sample remained at 51 candidates. The removal of these outliers did not significantly impact the intercept values, though, as the final fits were consistent within the uncertainties with the best-fit intercept before clipping was applied. The spatial positions of our final candidates in each galaxy are displayed in Figure \ref{fig:pos}. Color$-$magnitude diagrams with $3\sigma$ outliers marked are shown in Figure \ref{fig:cm}. The final best-fit light curve parameters and individual light curves of Cepheid candidates in each galaxy are available in the Appendix, with NGC\,4303 Cepheids in Section \ref{sec:app_a} and NGC\,1068 Cepheids in Section \ref{sec:app_b}.

\begin{deluxetable*}{lccccc}
    \tablewidth{0pt}
\tablecaption{Period$-$Luminosity Relationships}
\tablehead{
\colhead{Band} &
\colhead{$\alpha$} &
\colhead{$\beta$} &
\colhead{$\sigma$} &
\colhead{$\mu_{\rm rel \: LMC}$ (mag)} &
\colhead{$\mu$ (mag)} 
}

\startdata
\multicolumn{6}{c}{NGC\,4303} \\
\hline
F555W & $-2.76$ & $29.942 \pm 0.033$ & $0.525$ & $12.240 \pm 0.052$ & $30.717 \pm 0.058$ \\
F814W & $-2.96$ & $29.258 \pm 0.023$ & $0.326$ & $12.370 \pm 0.046$ & $30.847 \pm 0.053$ \\
$W_{I}$ & $-3.31$ & $28.491 \pm 0.021$ & $0.178$ & $12.556 \pm 0.021$ & $31.033 \pm 0.033$ \\
$W_{I}$ ($P > $ 19.5 days) & $-3.31$ & $28.491 \pm 0.021$ & $0.181$ & $12.556 \pm 0.021$ & $31.033 \pm 0.033$ \\
Metallicity-corrected $W_{I}$ ($P > $ 19.5 days) & \nodata & \nodata & \nodata & \nodata & $\mathbf{31.083 \pm 0.035}$ \\
\hline
\multicolumn{6}{c}{NGC\,1068} \\
\hline
F555W & $-2.76$ & $29.931 \pm 0.075$ & $1.058$ & $12.198 \pm 0.085$ & $30.675 \pm 0.089$ \\
F814W & $-2.96$ & $28.853 \pm 0.066$ & $0.900$ & $11.948 \pm 0.077$ & $30.425 \pm 0.082$ \\
$W_{I}$ & $-3.31$ & $27.575 \pm 0.086$ & $0.707$ & $11.640 \pm 0.088$ & $30.117 \pm 0.091$ \\
$W_{I}$ ($P > $ 25 days) & $-3.31$ & $27.503 \pm 0.100$ & $0.831$ & $11.568 \pm 0.100$ & $30.045 \pm 0.103$ \\
Metallicity-corrected $W_{I}$ ($P > $ 25 days) & \nodata & \nodata & \nodata & \nodata & $\mathbf{30.150 \pm 0.106}$ \\
\enddata
    \tablecomments{Slopes ($\alpha$) were held fixed to the values determined from LMC Cepheids by \cite{riess_large_2019}. LMC-relative distance moduli ($\mu_{\rm rel \: LMC}$) here have been corrected for extinction but not metallicity as discussed in Section \ref{sec:pl}. Distance moduli ($\mu$) were determined using an independently measured distance to the LMC by \cite{pietrzynski_distance_2019}. Our final adopted distance moduli, after correction for metallicity effects, are provided in bold text.}
    \label{tab:pl}
\end{deluxetable*}

Our sight-lines to both NGC\,4303 and NGC\,1068 are out of the plane of the Milky Way, and consequently the Galactic extinction is not severe. \cite{schlafly_measuring_2011} report Galactic extinction in the direction of NGC\,4303 to be $A_{V} = 0.064 \pm 0.040$ mag and $A_{I} = 0.034 \pm 0.040$ mag; extinction in the direction of NGC\,1068 is $A_{V} = 0.095 \pm 0.040$ mag and $A_{I} = 0.051 \pm 0.040$ mag. These values were subtracted from the intercepts in their respective filters. Due to the extinction-free nature of Wesenheit magnitudes, no correction was applied to $W_I$. Taking the LMC intercepts from \cite{riess_large_2019} with our extinction-corrected best-fit values, we calculated distance moduli to each galaxy relative to the LMC ($\mu_{\rm rel \: LMC}$). These values are listed in Table \ref{tab:pl}.

\begin{figure}[htb!]
    \centering
    \includegraphics[width=\columnwidth]{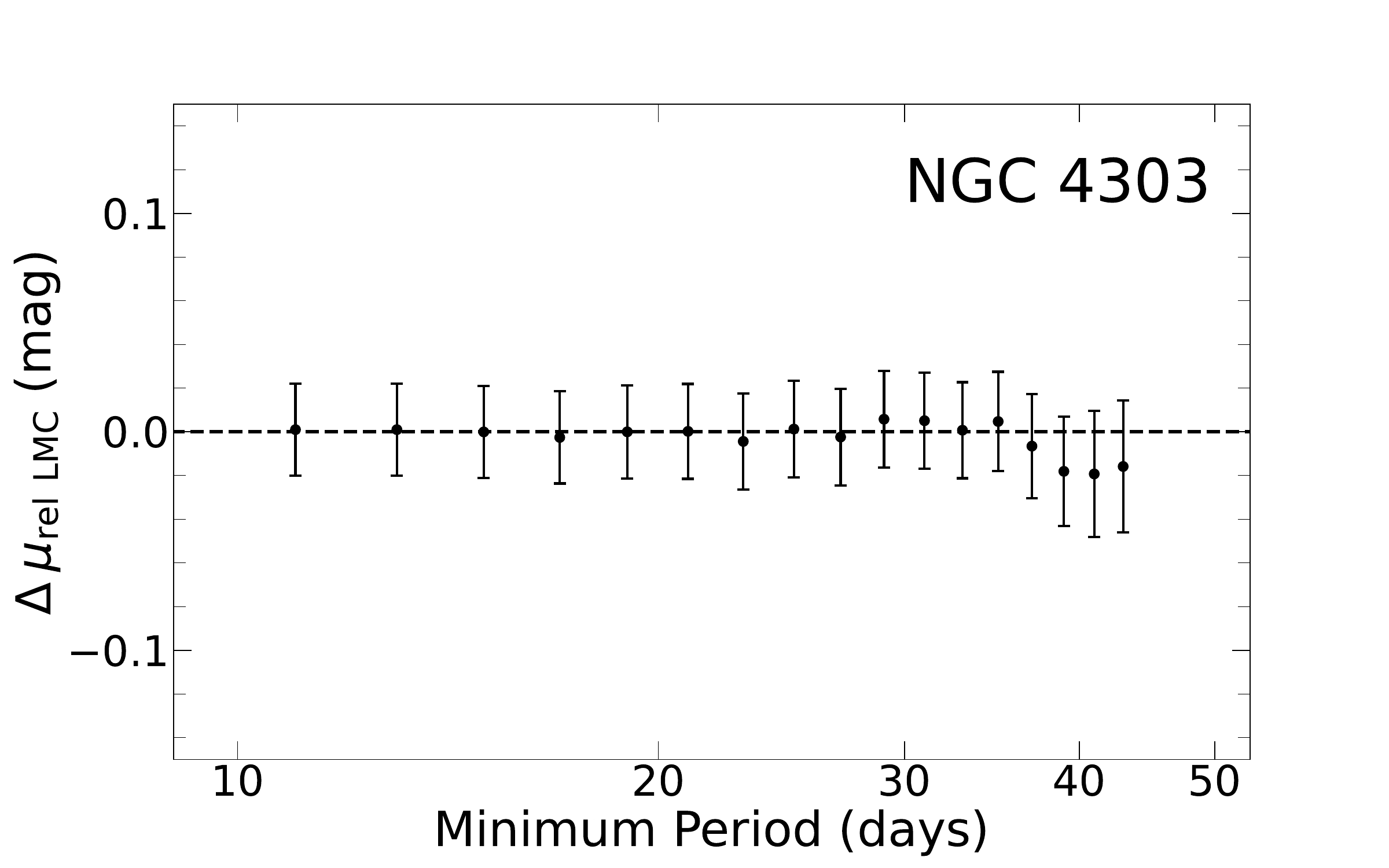}
    \includegraphics[width=\columnwidth]{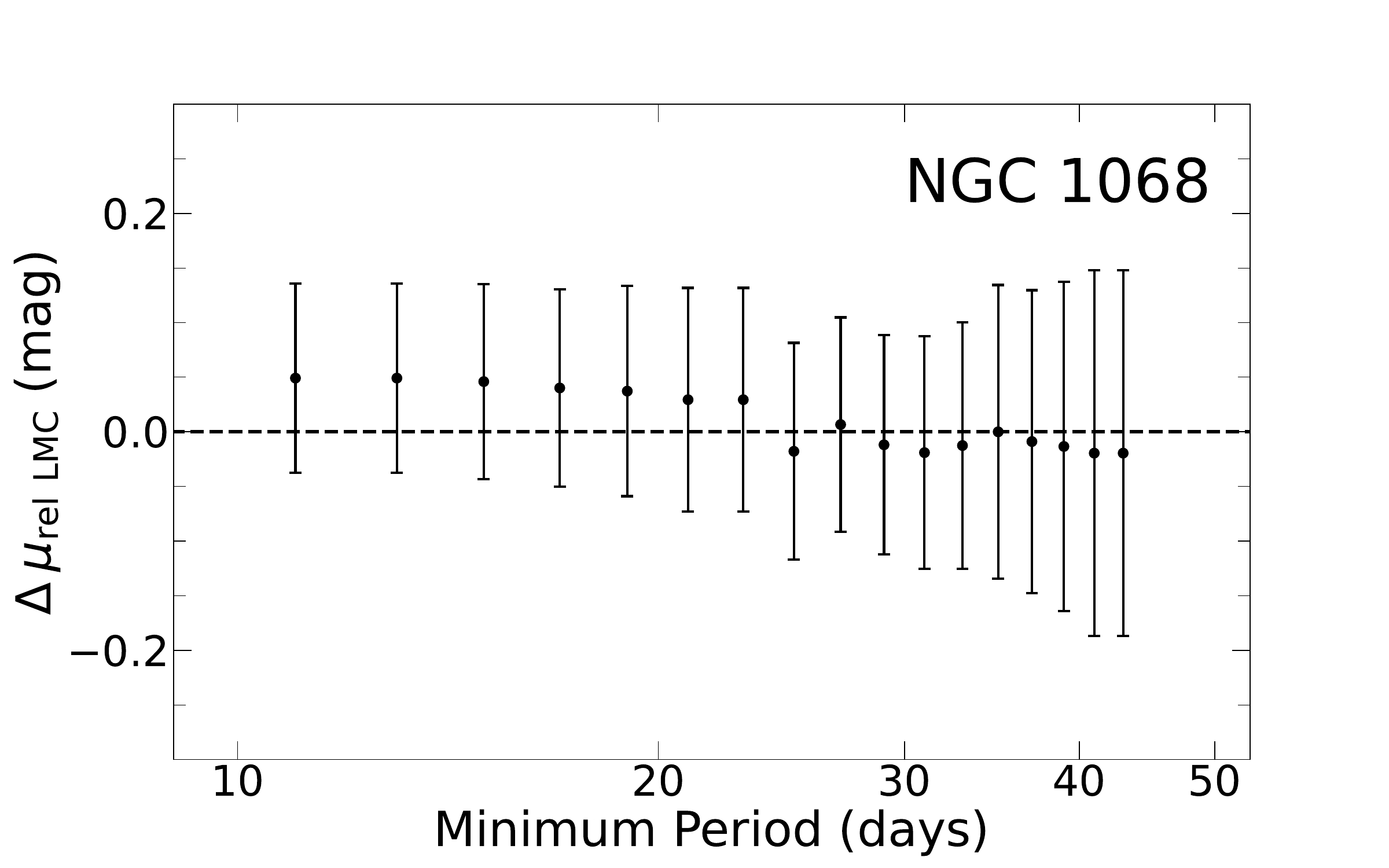}
    \caption{Dependence of $\mu_{\rm rel \: LMC}$ on the minimum period of Cepheid candidates in our final sample for NGC\,4303 (top) and NGC\,1068 (bottom). $\Delta \mu_{\rm rel \: LMC}$ is measured relative to the median value over all minimum period cuts. There is no significant evidence for incompleteness in NGC\,4303 towards shorter periods, though we do see potential incompleteness in NGC\,1068 at $P \lesssim 25$ days.}
    \label{fig:comp}
\end{figure}

We tested for incompleteness in our samples by recalculating the $W_{I}$ intercept over a series of minimum period cuts, shown in Figure \ref{fig:comp}. There is evidently no dependence on minimum period when determining the distance modulus for NGC\,4303, out to a minimum period cut of $\sim 50$ days. \cite{yuan_cepheid_2021} found a similar non-effect when testing for incompleteness in NGC 4051. However, there is a decrease in density around 19.5 days visible in Figure \ref{fig:PL}, so we chose to remove those candidates with periods $<19.5$ days. This left us with 124 Cepheid candidates, which was $\sim 95 \%$ of our $3\sigma$-clipped sample. In NGC\,1068, there is a hint of incompleteness in our sample at shorter periods. Based on Figure \ref{fig:comp}, we cut Cepheid candidates with $P < 25$ days from our sample. This left us with 38 Cepheid candidates for our distance determination, or $\sim 75 \%$ of our $3\sigma$-clipped sample for NGC\,1068.

\cite{pietrzynski_distance_2019} determined a distance modulus to the LMC that is precise to $1\%$ ($\mu_{\rm LMC} = 18.477 \pm 0.026$ mag); this value is currently the most precise available, and it is derived independently using eclipsing binaries. Using this value, we obtained distance moduli in each band to NGC\,4303 (see Table \ref{tab:pl}). The distance modulus using the $95\%$ sample of $W_{I}$ magnitudes results in a final distance determination of $D=16.09 \pm 0.25$ Mpc. The distance moduli we obtained for NGC\,1068 are also listed in Table \ref{tab:pl}. Our $75\%$ sample of $W_I$ magnitudes yields a distance to NGC\,1068 of $D = 10.21 \pm 0.49$ Mpc.

\subsection{Metallicity Dependence} \label{sec:metal}

The effects of metallicity on the period$-$luminosity relationship have been widely studied and hotly debated. For several decades, work to constrain $\gamma$, the correlation factor between the period$-$luminosity relationship zeropoint and metallicity, has yielded inconsistent results. Some authors have proposed that metal-rich Cepheids are systematically fainter (e.g., \citealt{romaniello_influence_2008, bono_cepheids_2008}), 
though most current debates center around whether the period$-$luminosity relationship displays a negative ($\gamma < 0 \rm \ mag \ dex^{-1}$) or null correlation ($\gamma \approx 0 \rm \ mag \ dex^{-1}$) with metallicity. Of the studies which suggest no significant correlation (e.g., \citealt{udalski_optical_2001, freedman_two_2011}), \cite{madore_systematics_2024} currently provide the largest sample in their study, with further support from \cite{madore_chicagocarnegie_2025}. They compare the zeropoints for 28 nearby ($D \leq 12.5$ Mpc) galaxies with distances derived from both Cepheids and the TRGB method, directly updating the work done by \cite{sakai_effect_2004}, which found a negative metallicity correlation. Using updated TRGB distances and a slightly larger sample, they find no significant dependence of the reddening-free Wesenheit period$-$luminosity relationship on metallicity (though see \citealt{breuval_small_2024} for possible limitations of the study). 

Other studies suggest that metal-rich Cepheids are brighter than metal-poor ones (e.g., \citealt{freedman_empirical_1990, kennicutt_hubble_1998, sakai_effect_2004, breuval_influence_2021}). Among the authors who suggest a negative correlation of the period$-$luminosity relationship with metallicity, many find a value of $\gamma \approx -0.25 \rm \: mag \: dex^{-1}$. Recently, \cite{breuval_small_2024} re-determined the calibration of the metallicity effect in the \textit{HST} WFC3 photometric system. Using samples of Cepheids with identical-system photometry in the SMC, LMC, and Milky Way galaxies, they find a correlation of $\gamma = -0.264 \pm 0.058  \rm \ mag \ dex^{-1}$ using $W_{I}$ magnitudes.

\subsubsection{NGC\,4303} \label{sec:met_n4303}

In addition to the disagreement regarding the effects of metallicity on the Cepheid period$-$luminosity relationship, there are disagreements in the literature surrounding metallicity gradient values for NGC\,4303. \cite{martin_oxygen_1992} produce an early calculation of $12 + \rm log([O/H]) = 9.38$ at the galactic center with a radial gradient of $-0.076 \pm 0.006 \ \rm dex \ kpc^{-1}$ ($-9.8 \pm 0.8 \times 10^{-7} \ \rm dex \ arcsec^{-1}$, from their assumed distance of 15.2 Mpc) based on line ratios from 79 \ion{H}{2} regions and using the \cite{edmunds_composition_1984} empirical relationship between O/H and [\ion{O}{3}]/H$\beta$. This relationship was calibrated using [\ion{O}{3}] ($\lambda4959+\lambda5007$) and H$\beta$ line measurements in the nuclear regions of NGC\,2997 and NGC\,7793, plus ten additional \ion{H}{2} regions between the two galaxies. The revision of solar composition by \cite{asplund_solar_2005} led to revisions of physical inputs for standard solar models, and this may affect older metallicity determinations. Recent work to update the metallicity gradient has been accomplished by \cite{groves_phangsmuse_2023}, who produced metallicity gradients for PHANGS galaxies using MUSE observations of nebular regions. They determined a central metallicity of $\rm 12+log([O/H]) = 8.613$ and a gradient of $-0.032 \pm 0.006 \ \rm dex \ r_{eff}^{-1}$ for NGC\,4303, where $r_{eff}$ is the effective radius containing half of the stellar mass of the galaxy. From their assumption of an $\rm r_{\rm eff}$ of $0.7'$, we converted the gradient from a change in metallicity over physical distance to a change in metallicity over angular distance, which yielded a gradient of $-7.6 \pm 1.4 \times 10^{-4} \ \rm dex \ arcsec^{-1}$. These metallicity calculations employed the empirical relationship by \cite{pilyugin_new_2016} which was calibrated through direct measurements of electron temperature using auroral lines of $>300$ \ion{H}{2} regions across a large sample of galaxies, and \cite{groves_phangsmuse_2023} considered it to be the most robust.

To investigate the potential effect of metallicity on our distance determination, we began with the findings of \cite{breuval_13_2023}. The correction to the distance modulus relative to the LMC is given by:
\begin{equation}
    \mu_{\rm rel \: LMC} = (\beta_{\rm NGC\,4303} - \beta_{\rm LMC}) + \Delta m
    \label{eq:metal_modulus}
\end{equation} where $\beta_{\rm NGC\,4303}$ is our best-fit intercept of the period$-$luminosity relationship using NGC\,4303 Cepheid candidates, and $\beta_{\rm LMC}$ is the intercept for the LMC period$-$luminosity fit as reported by \cite{riess_large_2019}. The correction $\Delta m$ accounts for the difference in metallicity between LMC Cepheids and NGC 4303 Cepheids:
\begin{equation}
    \Delta m = -\gamma ([\rm O/H]_{NGC\,4303} - [O/H]_{LMC})
    \label{eq:metal_calibration}
\end{equation} where $[\rm O/H]_{NGC\,4303}$ and $[\rm O/H]_{LMC}$ are the average metallicities of Cepheids in the two galaxies, and $\gamma = -0.264 \pm 0.058  \rm \ mag \ dex^{-1}$ \citep{breuval_small_2024}. We adopted $[\rm O/H]_{\rm LMC}=-0.32 \pm 0.01$ from \cite{romaniello_iron_2022}, relative to a solar abundance of $12 + \rm log([O/H]) = 8.69$ \citep{asplund_chemical_2009}. We used the deprojected angular separations of each Cepheid relative to the center of the galaxy with the gradient by \cite{groves_phangsmuse_2023} to determine metallicities at each location. The [O/H] metallicities in our sample of Cepheids candidates ranged from $-0.15$ to $-0.09$ dex (relative to solar), with a median of $-0.13 \pm 0.01$ dex.

We calculated the magnitude correction due to metallicity in two ways: first, by adopting a median metallicity for all Cepheid candidates after fitting the period$-$luminosity relationship and then determining a correction using Equation \ref{eq:metal_calibration}; and second, by using individual Cepheid metallicities with Equation \ref{eq:metal_calibration} to correct the $W_I$ magnitudes before the fitting and $3\sigma$-clipping procedures. The second method produced an identical $3\sigma$-clipped sample, and both methods yielded the same magnitude correction of $\Delta m = 0.050 \pm 0.012$ mag.

Using Equation \ref{eq:metal_modulus}, we applied this correction of $\Delta m = 0.050 \pm 0.012$ mag to our $95\%$ sample of $W_{I}$ magnitudes, finding final values of $\mu_{W_{I \rm (P\,>\,19.5 \: days)}} = 31.083 \pm 0.035$ mag and $D = 16.47 \pm 0.27$ Mpc. The inclusion of the $\Delta m$ correction added 0.38 Mpc to our distance constraint.

\subsubsection{NGC\,1068} \label{sec:met_n1068}

Recent studies have found a slight positive metallicity gradient in NGC\,1068 (e.g., \citealt{armah_spatially_2024,amiri_negative_2025}). We adopt the metallicity gradient from \cite{amiri_negative_2025}, which makes use of MUSE data from the MAGNUM survey \citep{mingozzi_m_magnum_2019, venturi_g_magnum_2021}. Using the empirical relationship of \cite{pettini_o_2004}, they determined a central metallicity of $\rm 12+log([O/H]) = 8.76$ and a gradient of $2.1 \pm 1.4 \times 10^{-5} \ \rm dex \ pc^{-1}$ for NGC\,1068. These authors assumed a distance of $D = 12.5$ Mpc, so we converted the gradient to $1.3 \pm 0.8 \times 10^{-4} \ \rm dex \ arcsec^{-1}$. The empirical relationship of \cite{pettini_o_2004} was derived from nebular [\ion{O}{3}] and [\ion{N}{2}] lines, and it was calibrated with a sample of 137 extragalactic \ion{H}{2} regions. All but six of these \ion{H}{2} regions had oxygen abundances determined from electron temperature. 

Following the same procedures as for NGC 4303, we first determined the expected metallicity of each Cepheid candidate by using the gradient value and the deprojected angular separations from the galactic center. In NGC\,1068, the [O/H] metallicities of our Cepheid candidates spanned $0.073-0.085$ dex (relative to solar), with a median of $0.077$ dex. 

Using Equation \ref{eq:metal_calibration}, we determined both individual corrections and a median correction for our sample of Cepheid candidates. These corrections are in very good agreement, and we adopted the median metallicity correction to the $W_I$ magnitudes of $\Delta m = 0.105 \pm 0.023$ mag.

Finally, we used Equation \ref{eq:metal_modulus} to correct our $W_I$ distance modulus, where $\Delta m = 0.105 \pm 0.023$ mag. Our metallicity-corrected $75\%$ sample of $W_I$ magnitudes yielded a distance modulus of $\mu_{W_{I \rm (P\,>\,25 \: days)}} = 30.150 \pm 0.106$ mag and a final distance determination of $D = 10.72 \pm 0.52$ Mpc. The inclusion of the $\Delta m$ correction added 0.51 Mpc to our distance constraint.

\section{Discussion} \label{sec:disc}

\subsection{Previous Distance Measurements}

\begin{figure*}[htb!]
    \centering
    \includegraphics[width=\columnwidth]{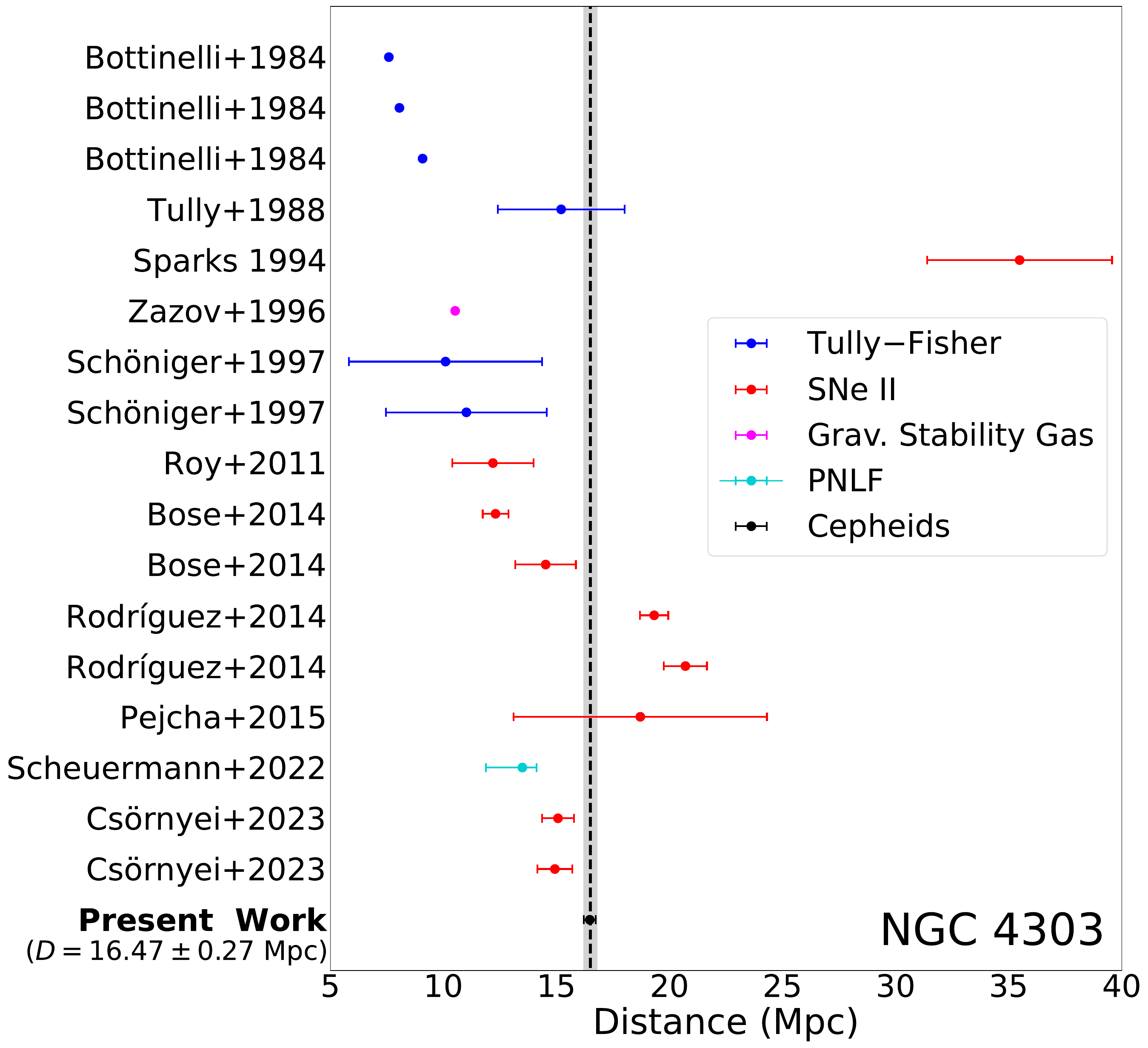}
    \includegraphics[width=\columnwidth]{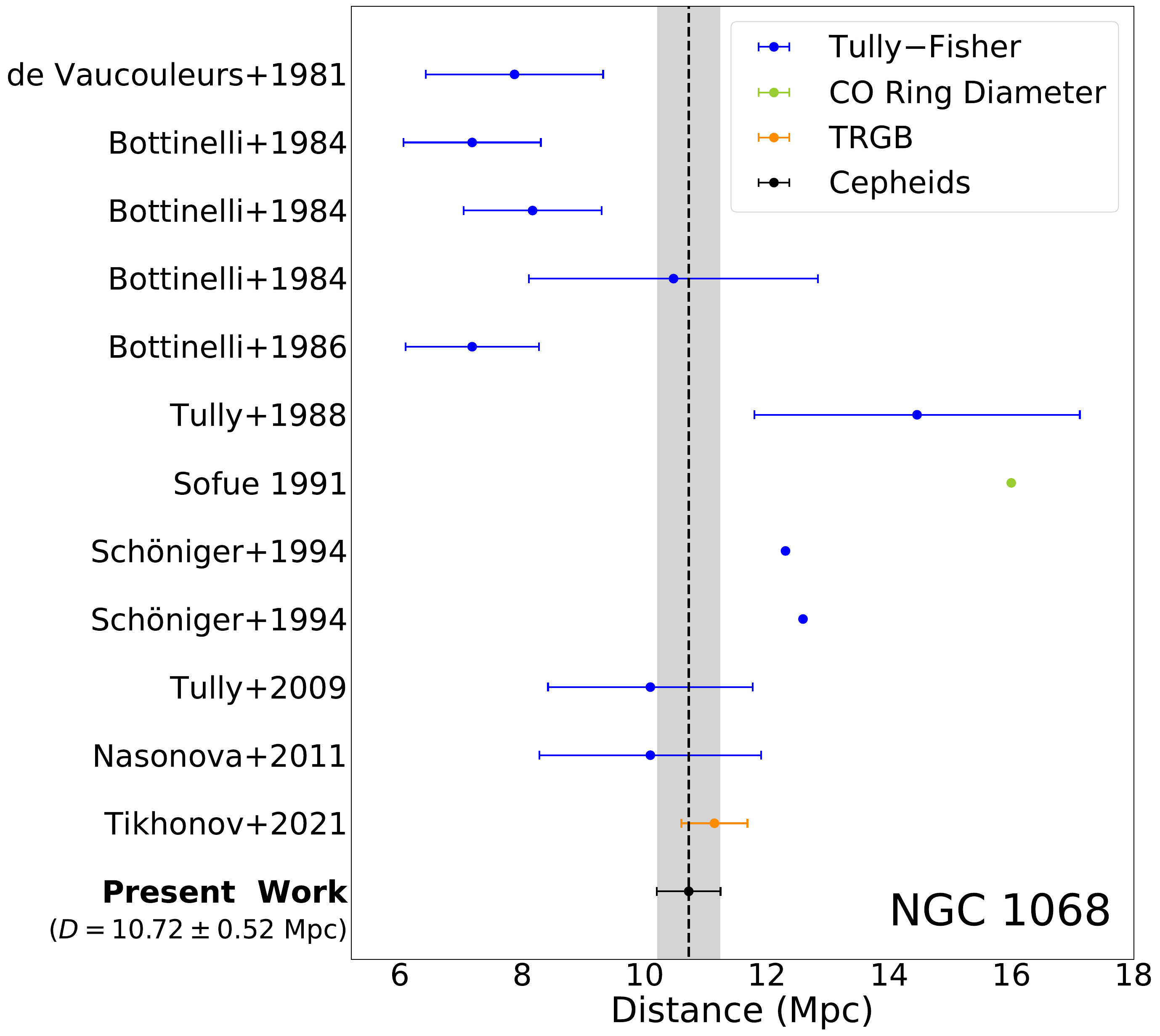}
    \caption{Comparison of previous distance measurements against our Cepheid-based distances for NGC\,4303 (left) and NGC\,1068 (right). Distance measurements are colored by their corresponding method. Multiple entries by the same authors coincide with multiple distance measurements reported in a single study. The vertical dashed lines represents our distance measurements, and the gray regions delimit our uncertainties.}
    \label{fig:dist_comp}
\end{figure*}

Redshift-independent distance measurements for NGC\,4303 span a wide range of values (see Figure \ref{fig:dist_comp}), from $7.59$ Mpc using the Tully$-$Fisher method \citep{bottinelli_hi_1984}, to $35.5$ Mpc using SNe II \citep{sparks_direct_1994}. One potential complication of Tully$-$Fisher distance measurements is the inclination angle \citep{tully_new_1977}. NGC\,4303 is relatively face-on ($i=25 \degree$, \citealt{schinnerer_toward_2002}), which increases the uncertainty in correcting the width of the \ion{H}{1} line for the effect of inclination. The most recent distance measurements from this method are 10.1 and 11.0 Mpc by \cite{schoniger_co_1997}. \cite{zasov_estimate_1996} provided a similar distance measurement of 10.5 Mpc based on gravitational stability of gas in disks. 

Distance measurements made using SNe II methods are mostly based on SN 2008in (\citealt{sparks_direct_1994} used SN 1926A, and \citealt{csornyei_consistency_2023} examined SN 2020jfo), and these measurements are not all in agreement. \cite{roy_sn_2011} used the standardized candle method to estimate a weighted mean distance of $13.19 \pm 1.09$ Mpc from mid-plateau apparent \textit{V-}band magnitude, \textit{I-}band magnitude, and photospheric velocity. \cite{rodriguez_photospheric_2014} find some of the largest distances by fitting photospheric magnitude diagrams, yielding distance measurements of $19.3-20.7$ Mpc for NGC\,4303.  \cite{bose_distance_2014} used two different models to derive distances with the expanding photosphere method \citep{hamuy_distance_2001,dessart_distance_2005}, and they find that these models differ by $30-50\%$ across their sample. \cite{pejcha_global_2015} used a variation on the expanding photosphere method, and their resulting distance for NGC\,4303 of $D = 18.73 \pm 5.57$ Mpc has a much greater uncertainty than other SNe II methods. However, recent progress has been made in standardizing SNe IIP (e.g., \citealt{2022MNRAS.514.4620D}); in particular, \cite{csornyei_consistency_2023} produced consistent results using two SNe IIP in NGC\,4303. They used the spectral-fitting method introduced by \cite{vogl_spectral_2020}, which takes a set of input parameters and predicts SNe IIP spectra. Applying this method to both SN 2008in and SN 2020jfo, they determined distances of $D = 15.06 \pm 0.71$ Mpc and $D = 14.95 \pm 0.78$ Mpc, respectively. These two distance measurements agree remarkably well, and the distance derived using SN 2008in is $2\sigma$ away from our distance measurement of $D = 16.47 \pm 0.27$ Mpc.

One distance often assumed for NGC\,4303 comes from a group average presented in \cite{kourkchi_galaxy_2017} using distances listed in \cite{tully_cosmicflows-3_2016} rather than a direct measurement of NGC\,4303. \cite{tully_cosmicflows-3_2016} reported NGC\,4303 as one of a 39-member group, where 18 members have reported distances that span the range $25.35-52.72$ Mpc with a group average distance reported as 33 Mpc. \cite{kourkchi_galaxy_2017} redefined galaxy groups within $3500 \rm \: km \: s^{-1}$ and adopted a weighted group average of $16.99 \pm 3.04$ Mpc for NGC\,4303, based on a new group that includes ten members with five distances reported spanning $16.00-26.30$ Mpc.

Recently, \cite{scheuermann_planetary_2022} determined a distance of $13.49^{+0.64}_{-1.60}$ Mpc for NGC\,4303 using PHANGS-MUSE data with the planetary nebula luminosity function (PNLF). Their findings highlight the potential effects of supernova remnant (SNR) contamination on PNLF distances. They classified 19 sources in NGC\,4303 as planetary nebulae (PNe) and an additional eight sources as SNR. The inclusion of seven out of eight SNR (with the brightest source rejected) improved the fit of the PNLF, however these authors note that the inclusion of all eight SNR yielded a distance that was nearly 3 Mpc closer. \cite{anand_distances_2021}, also using PHANGS-MUSE data, employed the TRGB method to find distances to several PHANGS galaxies. However, they were unsuccessful in determining a distance to NGC\,4303 using this method. They report that either an issue of depth below the TRGB or a lack or resolved sources in their chosen field limit their ability to constrain distances to some galaxies in their sample, including NGC\,4303. This conclusion is supported by our distance measurement of $D = 16.47 \pm 0.27$ Mpc, as a distance of $D \gtrsim 15$ Mpc would likely explain an insufficient depth in the PHANGS data for constraining the TRGB. 

In the case of NGC\,1068, the redshift is often quoted instead of a distance measurement ($z=0.003793$, \citealt{de_vaucouleurs_third_1991}), though redshift-independent distance measurements do exist. Reported distances span 7.18 Mpc using the Tully$-$Fisher method \citep{bottinelli_hi_1984} to 16.0 Mpc using CO emission from the central ring of molecular gas (\citealt{sofue_molecular_1991}, see Figure \ref{fig:dist_comp}). The majority of distance measurements for NGC\,1068 have been obtained using the Tully$-$Fisher method, though studies find a factor of $\sim 2$ spread in measurements using this method ($7.18 - 14.4$ Mpc, \citealt{bottinelli_hi_1984, bottinelli_malmquist_1986,tully_catalog_1988,schoniger_co_1994}). NGC\,1068 is moderately inclined ($i=40 \pm 3 \degree$, \citealt{brinks_hi_1997}), but it also has a more powerful AGN that could potentially impact Tully$-$Fisher results. The most recent distance measurement using the Tully$-$Fisher method is $10.1 \pm 2.0$ Mpc \citep{tully_extragalactic_2009, nasonova_hubble_2011}.

Other distances include $D = 11.14 \pm 0.54$ Mpc based on the Numerical Action Methods (NAM) of \cite{shaya_action_2017} and \cite{kourkchi_cosmicflows-3_2020}, though it is important to note that this distance is a model-based estimate rather than a measurement. Like NGC\,4303, a TRGB method distance measurement has also been attempted for NGC\,1068. \cite{tikhonov_trgb_2021} find a distance of $11.14 \pm 0.54$ Mpc using archival \textit{HST} data in the F606W and F814W filters. However, the fields of the observations in the two filters do not overlap, and these authors consequently adopt an average $V-I$ color for metal-poor red giants ($V-I = 1.4 \pm 0.1$ mag). Our Cepheid-based distance measurement of $D = 10.72 \pm 0.52$ Mpc agrees fairly well with this TRGB measurement.

\subsection{Peculiar Velocities and Environment}

One open question regarding NGC 4303 is its association (or lack thereof) with the Virgo cluster. This galaxy cluster is expansive on the sky, and it is composed of several substructures that are distinct in their mean recessional velocities (e.g., \citealt{binggeli_studies_1985}). \cite{cantiello_next_2024} recently produced a three-dimensional distribution of galaxies in the Virgo cluster using surface brightness fluctuation (SBF) distances. Our distance measurement to NGC\,4303 is closest to the median distance of cluster B \citep{binggeli_kinematics_1993}, though its position on the sky is significantly ($>2 \degree$) outside of the cluster B radius, as defined by \cite{boselli_galex_2014}. There are southern cloud-like substructures closer to the position of NGC\,4303 on the sky, and \cite{binggeli_kinematics_1993} placed NGC\,4303 as a member of the W cloud substructure. However, the average distances of these associations determined by \cite{cantiello_next_2024} are $22.7-29.0$ Mpc, which is located behind NGC\,4303. \cite{karachentsev_intense_2013} characterize the dimensions of the Southern Extension, which they describe as a sheet with a depth of $\sim 15$ Mpc and a median distance of $17 \pm 2$ Mpc. \cite{yoon_warm_2012} find evidence for a connective filament between the Southern Extension, W cloud, and M cloud. By examining Ly$\alpha$ absorption in and surrounding the Virgo cluster, they suggest the presence of a large-scale ($> 3.6$ Mpc) flow of warm gas. Based on our distance measurement and the location of NGC\,4303 on the sky, it seems most likely that it is associated with either the Southern Extension or this connective filament.

To further investigate the association of NGC\,4303 with the Virgo cluster, we determined the peculiar velocities affecting this galaxy. An \ion{H}{1} recessional velocity of $1566 \: \rm km \: s^{-1}$ is provided by \cite{haynes_arecibo_2018}. Adopting $H_0 = 73 \: \rm km \: s^{-1} \: Mpc^{-1}$ \citep{riess_comprehensive_2022} predicts a distance of $D = 21.5$ Mpc. Compared with our Cepheid-based distance, the redshift implies peculiar velocities of $\sim 300-400 \: \rm km \: s^{-1}$. A distance of $D = 16.47 \pm 0.27$ Mpc is similar to the average distances of both the Virgo cluster and the Southern Extension, though NGC\,4303 appears to be moving away at a fairly high rate.

We also determined the peculiar velocities affecting NGC\,1068. \cite{bottinelli_extragalactic_1990} report an \ion{H}{1} recessional velocity of $1137 \: \rm km \: s^{-1}$. Once again adopting $H_0 = 73 \: \rm km \: s^{-1} \: Mpc^{-1}$ \citep{riess_comprehensive_2022}, we obtain a distance of $D = 15.6$ Mpc. Similar to NGC\,4303, we found peculiar velocities of $\sim 300-400 \: \rm km \: s^{-1}$ affecting NGC\,1068. While NGC\,1068 is not associated with a nearby massive galaxy cluster like the Virgo or Fornax clusters, \cite{kourkchi_galaxy_2017} determined it to be a part of a nine-member group. Two of the nine members have distances reported ($D=19.41$ Mpc for NGC\,1055, $D=14.00$ Mpc for NGC\,1087). The peculiar velocities affecting NGC\,1068 may be related to its proximity to other massive galaxies.

\cite{tully_cosmicflows-2_2013} have shown that peculiar velocities for nearby galaxies can reach $\sim 500 \: \rm km \: s^{-1}$, and it is not unusual for galaxies in groups to have peculiar velocities on the order of several $100 \: \rm km \: s^{-1}$. As is evident from both galaxies in our sample, redshift alone is not sufficient to derive accurate distances to nearby AGN.

\subsection{Implications of New Distance Measurements}

Considering the effect of our distance measurement on the black hole mass, a Cepheid-based distance of $D=16.47 \pm 0.27$ Mpc for NGC\,4303 agrees fairly well with the distance adopted by \cite{pastorini_supermassive_2007} ($D = 16.1$ Mpc). Therefore, the central black hole mass they derive ($M_{\rm BH} = 5.0^{+0.9}_{-2.3} \times 10^6 \ M_{\odot}$) is consistent within the given uncertainties. In their studies on the starburst$-$AGN connection, \cite{riffel_sinfoni_2016} and \cite{dametto_sinfoni_2019} adopt the same distance of $D = 16.1$ Mpc. The size of the circumnuclear star-forming ring depends linearly on distance while the star formation rate (using \citealt{kennicutt_star_1998}) and the masses of hot molecular and ionized gas inside the ring depend on distance squared. The findings of \cite{riffel_sinfoni_2016} and \cite{dametto_sinfoni_2019} are changed by $< 5 \%$ by adopting our Cepheid-based distance. The luminosity of the nuclear region of NGC\,4303 is dominated by star formation activity with weak contribution from a low-luminosity AGN \citep{colina_detection_2002, jimenez-bailon_nuclear_2003}. The UV and X-ray luminosities determined by these authors ($\sim10^{39} \: \rm erg \: s^{-1}$) also assume a distance of $D = 16.1$ Mpc, so these values remain relatively unchanged as well. 

There are, however, several studies that have adopted quite different distances to NGC\,4303. \cite{davis_black_2019} examined the SMBH mass ($M_{BH}$)$-$bulge stellar mass ($M_{*,sph}$) scaling relation with a sample of spiral galaxies. They assumed a distance of $D = 12.3 \pm 0.6$ Mpc for NGC\,4303, which is a factor of $\sim 1.3$ times smaller than our distance of $D = 16.47 \pm 0.27$ Mpc. Taking the stellar mass $M_{*,sph}$ from a mass-to-light ratio, their $M_{*,sph}$ value for NGC\,4303 is $\sim 1.7$ times smaller than would be implied by our new distance. \cite{graham_appreciating_2023} presented an updated $M_{BH} - M_{*,sph}$ scaling relation with a larger sample, and in this study, a distance of $D = 19.3 \pm 0.6$ was assumed for NGC\,4303 ($\sim 1.2$ times greater than our distance). In this case, the $M_{*,sph}$ value for NGC\,4303 is $\sim 1.4$ times greater than what would be implied from our new distance measurement of $D = 16.47 \pm 0.27$ Mpc. An accurate distance for NGC\,4303 thus has implications beyond the individual galaxy properties because it has the potential to also impact broader scaling relations.

The many different distances that have been adopted for NGC\,1068 affect the results of several previous studies. Our Cepheid-based distance measurement of $D = 10.72 \pm 0.52$ Mpc for NGC\,1068 is quite a bit smaller than the Tully$-$Fisher measurement of $D = 14.4$ Mpc \citep{tully_catalog_1988}; the two differ by a factor of $\sim 1.3$. \cite{lodato_non-keplerian_2003} derive a black hole mass of $M_{\rm BH} = 8.0 \pm 0.3 \times 10^6 \ M_{\odot}$ for NGC\,1068, assuming $D = 14.4$ Mpc. Scaling linearly with distance, our Cepheid-based measurement implies a mass closer to $\sim 6 \times 10^6 \ M_{\odot}$. Several studies of the kinematics and structure of gas in the torus of NGC\,1068 (e.g., \citealt{garcia-burillo_molecular_2014, gamez_rosas_decoding_2025}) also assume a distance of 14.4 Mpc, which affects both the AGN luminosity and the reported scale of the outflows. \cite{garcia-burillo_molecular_2014} derived an AGN bolometric luminosity of $4.2^{+1.4}_{-1.1} \times 10^{44} \: \rm erg \: s^{-1}$, estimated from CLUMPY model fits to the nuclear SED of NGC\,1068. \cite{gamez_rosas_decoding_2025} assumed an AGN bolometric luminosity of $(0.4-4.7) \times 10^{45} \: \rm erg \: s^{-1}$, from \cite{gravity_collaboration_image_2020}. Our distance measurement of $D = 10.72 \pm 0.52$ Mpc implies an AGN luminosity that is nearly a factor of 2 smaller than the values reported by these authors.

Some authors have adopted a redshift rather than a distance measurement for NGC\,1068. \cite{spinoglio_far-infrared_2005} quoted a redshift of $z = 0.0038$ and corresponding distance of $D = 15.2$ Mpc (assuming $H_0 = 75 \: \rm km \: s^{-1} \: Mpc^{-1}$) in their study of the far-infrared spectrum of NGC\,1068. This distance is $\sim 1.4$ times larger than our distance measurement, and inaccurate line luminosities yield inaccuracies in the input physical parameters of models that are fit to the SED. In their study of the X-ray spectrum, \cite{kamenetzky_dense_2011} assumed $H_0 = 70 \: \rm km \: s^{-1} \: Mpc^{-1}$ for the same redshift ($z = 0.0038$), and this implies a distance of $D = 16.2$ Mpc. Their angular size scale is $\sim 1.5$ larger than what is implied by our distance measurement of $D = 10.72 \pm 0.52$. The importance of an accurate distance measurement for NGC\,1068 is evident in all wavelength regimes.

\section{Summary} \label{sec:sum}

We obtained multi-epoch HST imaging to identify and characterize Cepheid candidates in two canonical AGN host galaxies\,---\,NGC\,4303 and NGC\,1068. For each sample of Cepheid candidates, we fit the extinction-free $W_{I}$ period$-$luminosity relationship to derive distance moduli. Carefully considering the potential effects of metallicity, we find $\mu = 31.083 \pm 0.035$ mag for NGC\,4303 and $\mu = 30.150 \pm 0.106$ mag for NGC\,1068. These correspond to distances of $D= 16.47 \pm 0.27$ Mpc and $D = 10.72 \pm 0.52$ Mpc, respectively.

Our Cepheid-based distance to NGC\,4303 agrees fairly well with the distance commonly used to determine its black hole mass and nuclear luminosity ($D = 16.1$ Mpc). Therefore, these characteristics are relatively unaffected by our new distance measurement. However, other work that adopted different distances will be affected, including studies of scaling relations such as the $M_{BH} - M_{*,sph}$ relationship. Our distance to NGC\,1068 differs from the often adopted $D = 14.4$ Mpc by a factor of $\sim 1.3$. This implies a factor of $\sim 1.3$ decrease in central black hole mass and a factor of $\sim 1.7$ decrease in AGN luminosity measurements. Our distance measurements for both NGC\,4303 and NGC\,1068 imply peculiar velocities of $\sim 300-400 \: \rm km \: s^{-1}$, which demonstrates that redshift alone is insufficient for estimating their distances.

\begin{acknowledgments}
We thank the referee for their helpful suggestions. We thank Judy Schmidt for her production of the images displayed in Figure \ref{fig:gals}.  M.\ Markham thanks Eugene Vasiliev for his constructive comments. We thank D. Michael Crenshaw and Monica Valluri for their contributions to the proposal. M.\ Markham and M.C.\ Bentz gratefully acknowledge support through program HST GO-17165, which was provided by NASA through a grant from the Space Telescope Science Institute, which is operated by the Association of Universities for Research in Astronomy, Inc., under NASA contract NAS 5–26555. M.\ Vestergaard gratefully acknowledges financial support from the Independent Research Fund Denmark via grant number DFF 3103-00146 and from the Carlsberg Foundation (grant CF23-0417).
\end{acknowledgments}

\begin{contribution}
M.\ Markham drizzled the observations, carried out the analysis, and wrote the manuscript. M.C.\ Bentz led the proposal and Phase 2 planning and oversaw the project. L.\ Ferrarese, C.A.\ Onken, and M.\ Vestergaard provided scientific expertise. All authors contributed to the presentation of the results.
\end{contribution}

\facility{HST (WFC3)}

\software{AstroDrizzle \citep{fruchter_betadrizzle_2010}, DAOPHOT \citep{stetson_daophot_1987}}

\appendix

\section{NGC\,4303} \label{sec:app_a}

\begin{figure*}[!ht]
    \centering
    \includegraphics[width=\columnwidth]{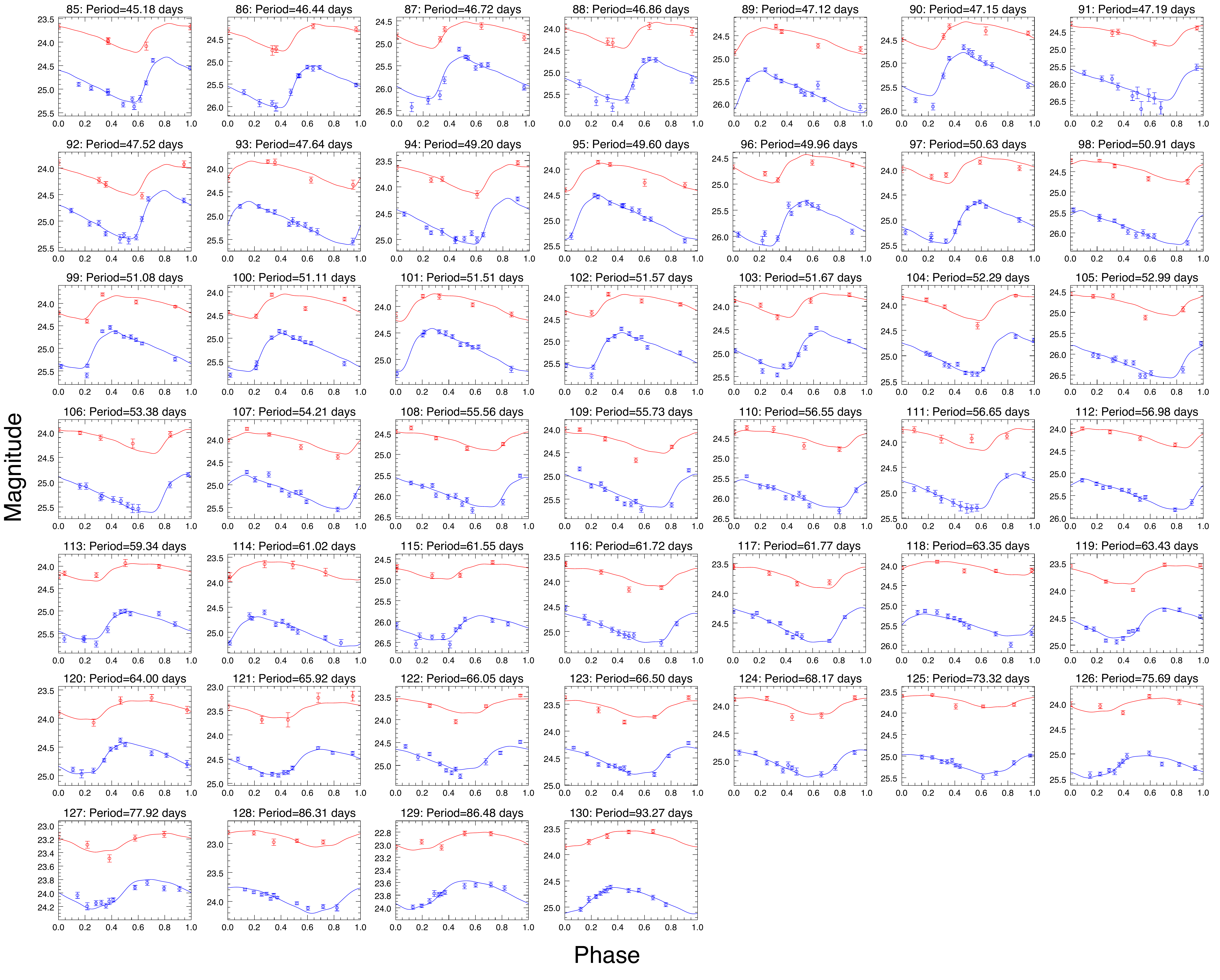}
    \caption{Light curves for 130 final Cepheid candidates in NGC\,4303. The data points show the measurements in F555W (blue) and F814W (red), while the solid lines represent the best Cepheid template fits to the data.}
    \label{fig:n4303_lcs}
\end{figure*}

The Cepheid candidates in Table \ref{tab:n4303_ceph} and Figure \ref{fig:n4303_lcs} represent our final $3\sigma$-clipped sample in NGC\,4303. Sources with periods below our visually-determined point of completeness ($P>19.5$ days) are included here.

\startlongtable
\begin{deluxetable*}{rccccccc}
    \tablecaption{Best-fit Light Curve Parameters for Final Cepheid Candidates in NGC\,4303}

\label{tab:n4303_ceph}

\tablehead{
\colhead{ID} &
\colhead{RA} &
\colhead{Dec} &
\colhead{$m_{\rm F555W}$} &
\colhead{$m_{\rm F814W}$} &
\colhead{$P$} &
\colhead{Phase} &
\colhead{$\chi^2 / \nu$} \\ 
\colhead{} &
\colhead{(hh:mm:ss)} &
\colhead{(dd:mm:ss)} &
\colhead{(mag)} &
\colhead{(mag)} &
\colhead{(days)} &
\colhead{(days)} &
\colhead{}
}

\startdata
1 & 12:21:53.263 & +04:27:03.486 & 25.98 ± 0.08 & 26.76 ± 0.04 & 13.17 ± 0.17 & 12.00 ± 0.32 & 1.35 \\
2 & 12:21:53.468 & +04:27:16.826 & 25.39 ± 0.05 & 26.28 ± 0.04 & 15.83 ± 0.21 & 4.56 ± 0.37 & 1.73 \\
3 & 12:21:51.230 & +04:27:21.124 & 25.85 ± 0.05 & 26.95 ± 0.03 & 15.82 ± 0.21 & 7.26 ± 0.47 & 1.28 \\
4 & 12:21:56.212 & +04:27:10.250 & 25.62 ± 0.04 & 26.17 ± 0.02 & 16.66 ± 0.18 & 6.31 ± 0.40 & 1.13 \\
5 & 12:21:54.775 & +04:27:35.278 & 25.18 ± 0.05 & 26.05 ± 0.03 & 17.42 ± 0.36 & 11.70 ± 0.39 & 1.15 \\
6 & 12:21:52.141 & +04:28:19.510 & 25.43 ± 0.04 & 26.35 ± 0.02 & 18.68 ± 0.10 & 5.21 ± 0.23 & 0.77 \\
7 & 12:21:57.700 & +04:28:54.315 & 25.10 ± 0.05 & 25.81 ± 0.03 & 20.01 ± 0.48 & 8.70 ± 0.50 & 2.62 \\
8 & 12:21:56.455 & +04:28:50.334 & 25.16 ± 0.06 & 26.14 ± 0.04 & 20.38 ± 0.14 & 9.64 ± 0.25 & 2.09 \\
9 & 12:21:57.615 & +04:27:56.948 & 25.10 ± 0.04 & 25.76 ± 0.03 & 20.69 ± 0.45 & 14.02 ± 0.63 & 1.66 \\
10 & 12:21:51.716 & +04:27:10.241 & 25.50 ± 0.06 & 26.46 ± 0.05 & 21.06 ± 0.69 & 15.49 ± 0.22 & 2.67 \\
11 & 12:21:58.877 & +04:27:49.639 & 25.75 ± 0.04 & 26.70 ± 0.02 & 21.28 ± 0.35 & 1.45 ± 0.67 & 0.72 \\
12 & 12:21:50.588 & +04:28:54.812 & 25.31 ± 0.07 & 26.27 ± 0.04 & 22.14 ± 0.37 & 3.45 ± 1.00 & 3.22 \\
13 & 12:21:51.592 & +04:28:06.725 & 24.95 ± 0.05 & 25.92 ± 0.03 & 22.49 ± 0.20 & 17.14 ± 0.28 & 1.71 \\
14 & 12:21:53.282 & +04:27:36.688 & 24.90 ± 0.05 & 25.75 ± 0.06 & 22.72 ± 0.53 & 14.01 ± 0.83 & 2.21 \\
15 & 12:21:51.569 & +04:29:18.385 & 25.50 ± 0.05 & 26.49 ± 0.03 & 22.82 ± 0.23 & 22.41 ± 0.26 & 2.06 \\
16 & 12:21:50.933 & +04:27:27.725 & 25.29 ± 0.07 & 26.72 ± 0.05 & 24.11 ± 0.82 & 8.50 ± 1.09 & 2.69 \\
17 & 12:21:50.895 & +04:29:12.196 & 24.76 ± 0.04 & 25.66 ± 0.02 & 24.34 ± 0.33 & 7.59 ± 0.36 & 1.34 \\
18 & 12:21:54.962 & +04:29:13.140 & 25.08 ± 0.04 & 25.78 ± 0.03 & 25.20 ± 0.39 & 11.97 ± 0.71 & 0.67 \\
19 & 12:21:54.752 & +04:27:31.321 & 25.32 ± 0.05 & 26.24 ± 0.05 & 26.14 ± 0.81 & 19.39 ± 1.07 & 2.24 \\
20 & 12:21:56.100 & +04:27:07.301 & 24.64 ± 0.03 & 25.48 ± 0.02 & 26.80 ± 0.40 & 5.63 ± 0.43 & 1.19 \\
21 & 12:21:50.748 & +04:28:48.214 & 24.90 ± 0.04 & 26.17 ± 0.03 & 27.12 ± 0.91 & 22.27 ± 0.30 & 1.73 \\
22 & 12:21:54.770 & +04:27:23.451 & 25.11 ± 0.07 & 26.03 ± 0.04 & 27.24 ± 0.62 & 25.26 ± 0.36 & 4.51 \\
23 & 12:21:50.374 & +04:29:07.801 & 24.90 ± 0.05 & 25.62 ± 0.03 & 27.54 ± 0.62 & 17.89 ± 0.52 & 4.36 \\
24 & 12:21:50.941 & +04:27:29.845 & 24.66 ± 0.05 & 25.73 ± 0.02 & 28.71 ± 0.80 & 1.95 ± 0.83 & 2.21 \\
25 & 12:21:56.052 & +04:28:07.481 & 24.43 ± 0.04 & 25.33 ± 0.03 & 28.91 ± 0.63 & 13.99 ± 0.32 & 2.22 \\
26 & 12:21:59.113 & +04:28:36.523 & 24.75 ± 0.05 & 25.53 ± 0.05 & 28.95 ± 0.74 & 12.96 ± 0.63 & 3.25 \\
27 & 12:21:57.359 & +04:29:23.384 & 24.69 ± 0.04 & 25.76 ± 0.03 & 29.22 ± 0.57 & 4.14 ± 0.97 & 2.10 \\
28 & 12:21:52.683 & +04:27:02.249 & 25.26 ± 0.03 & 26.31 ± 0.02 & 30.17 ± 0.22 & 11.33 ± 0.25 & 1.52 \\
29 & 12:21:52.047 & +04:27:28.152 & 24.73 ± 0.06 & 25.92 ± 0.05 & 31.36 ± 1.02 & 8.08 ± 0.88 & 4.68 \\
30 & 12:21:53.750 & +04:27:41.434 & 24.42 ± 0.05 & 25.54 ± 0.03 & 31.52 ± 0.50 & 12.16 ± 0.46 & 1.91 \\
31 & 12:21:50.549 & +04:28:47.959 & 25.41 ± 0.06 & 26.50 ± 0.04 & 31.59 ± 0.84 & 21.56 ± 0.29 & 2.95 \\
32 & 12:21:53.049 & +04:29:12.018 & 25.06 ± 0.02 & 26.14 ± 0.01 & 31.61 ± 0.23 & 21.20 ± 0.11 & 0.60 \\
33 & 12:21:56.518 & +04:28:27.730 & 24.41 ± 0.04 & 25.11 ± 0.02 & 32.35 ± 0.38 & 32.34 ± 0.25 & 1.34 \\
34 & 12:21:50.573 & +04:28:18.871 & 25.18 ± 0.04 & 26.27 ± 0.04 & 32.48 ± 0.69 & 7.25 ± 1.18 & 1.20 \\
35 & 12:21:54.057 & +04:29:21.817 & 24.89 ± 0.03 & 25.89 ± 0.03 & 32.75 ± 0.42 & 30.71 ± 0.20 & 1.04 \\
36 & 12:21:50.121 & +04:28:52.686 & 24.98 ± 0.03 & 26.06 ± 0.03 & 33.31 ± 0.75 & 21.93 ± 0.25 & 1.80 \\
37 & 12:21:56.631 & +04:27:31.086 & 24.49 ± 0.08 & 25.68 ± 0.08 & 34.03 ± 2.02 & 13.56 ± 2.51 & 7.42 \\
38 & 12:21:50.106 & +04:29:07.606 & 24.90 ± 0.04 & 26.10 ± 0.03 & 34.52 ± 0.92 & 21.23 ± 0.25 & 2.15 \\
39 & 12:21:56.277 & +04:28:24.736 & 24.33 ± 0.08 & 25.43 ± 0.05 & 34.59 ± 0.61 & 0.22 ± 0.81 & 4.40 \\
40 & 12:21:52.636 & +04:28:52.771 & 24.61 ± 0.05 & 25.60 ± 0.04 & 34.94 ± 1.26 & 12.85 ± 0.83 & 3.96 \\
41 & 12:21:50.683 & +04:28:53.698 & 24.67 ± 0.06 & 25.85 ± 0.04 & 34.97 ± 0.83 & 17.93 ± 0.62 & 5.92 \\
42 & 12:21:57.362 & +04:28:51.644 & 24.77 ± 0.05 & 25.74 ± 0.03 & 34.98 ± 0.66 & 11.27 ± 0.23 & 2.41 \\
43 & 12:21:59.140 & +04:28:00.798 & 25.10 ± 0.07 & 26.04 ± 0.06 & 35.14 ± 1.51 & 12.55 ± 1.57 & 5.22 \\
44 & 12:21:53.705 & +04:27:43.014 & 24.31 ± 0.06 & 25.11 ± 0.04 & 35.19 ± 1.20 & 12.36 ± 1.21 & 3.72 \\
45 & 12:21:56.688 & +04:29:32.895 & 24.73 ± 0.05 & 25.68 ± 0.04 & 35.68 ± 0.78 & 1.36 ± 0.68 & 2.12 \\
46 & 12:21:59.724 & +04:28:36.650 & 24.63 ± 0.05 & 25.63 ± 0.04 & 35.83 ± 1.52 & 19.85 ± 0.74 & 5.29 \\
47 & 12:21:53.441 & +04:29:03.488 & 24.70 ± 0.04 & 25.45 ± 0.02 & 35.89 ± 0.45 & 10.10 ± 0.24 & 2.29 \\
48 & 12:21:56.183 & +04:28:28.921 & 24.57 ± 0.05 & 25.48 ± 0.03 & 35.94 ± 0.55 & 0.96 ± 0.44 & 3.04 \\
49 & 12:21:55.016 & +04:27:29.317 & 24.54 ± 0.03 & 25.37 ± 0.02 & 36.17 ± 0.36 & 0.05 ± 0.32 & 1.65 \\
50 & 12:21:56.566 & +04:28:30.630 & 24.81 ± 0.05 & 25.89 ± 0.04 & 36.21 ± 1.13 & 18.16 ± 0.67 & 3.16 \\
51 & 12:21:53.763 & +04:29:33.005 & 24.33 ± 0.04 & 25.36 ± 0.03 & 36.61 ± 0.55 & 36.50 ± 0.48 & 4.71 \\
52 & 12:21:49.586 & +04:28:48.503 & 24.60 ± 0.05 & 25.60 ± 0.04 & 36.80 ± 1.90 & 28.66 ± 0.29 & 5.08 \\
53 & 12:21:50.605 & +04:28:45.944 & 24.69 ± 0.04 & 25.67 ± 0.03 & 37.57 ± 1.22 & 9.83 ± 0.35 & 2.88 \\
54 & 12:21:59.140 & +04:27:52.247 & 24.81 ± 0.05 & 25.78 ± 0.03 & 38.10 ± 1.12 & 10.38 ± 0.37 & 4.12 \\
55 & 12:21:55.279 & +04:27:05.601 & 24.75 ± 0.03 & 25.81 ± 0.02 & 38.15 ± 0.51 & 33.86 ± 0.36 & 1.36 \\
56 & 12:21:55.686 & +04:27:04.628 & 24.44 ± 0.04 & 25.32 ± 0.03 & 38.54 ± 0.66 & 10.97 ± 0.61 & 5.09 \\
57 & 12:21:55.482 & +04:29:17.255 & 24.45 ± 0.04 & 25.53 ± 0.02 & 38.64 ± 0.39 & 22.25 ± 0.25 & 1.77 \\
58 & 12:21:54.900 & +04:29:19.709 & 24.46 ± 0.03 & 25.32 ± 0.02 & 38.69 ± 0.77 & 2.34 ± 0.55 & 1.46 \\
59 & 12:21:52.628 & +04:29:12.704 & 24.68 ± 0.03 & 25.71 ± 0.02 & 39.00 ± 0.56 & 12.16 ± 0.43 & 3.06 \\
60 & 12:21:56.313 & +04:28:52.842 & 24.54 ± 0.02 & 25.70 ± 0.02 & 39.02 ± 0.43 & 23.39 ± 0.32 & 0.79 \\
61 & 12:21:54.548 & +04:27:41.062 & 24.82 ± 0.06 & 25.86 ± 0.04 & 39.15 ± 1.74 & 18.22 ± 0.57 & 3.57 \\
62 & 12:21:53.587 & +04:29:21.745 & 23.96 ± 0.06 & 24.76 ± 0.03 & 39.20 ± 0.42 & 30.37 ± 0.28 & 3.38 \\
63 & 12:21:54.507 & +04:27:01.090 & 24.81 ± 0.04 & 25.94 ± 0.03 & 39.32 ± 0.58 & 11.33 ± 0.40 & 4.37 \\
64 & 12:21:52.913 & +04:28:14.473 & 25.11 ± 0.07 & 26.64 ± 0.06 & 39.41 ± 1.07 & 11.44 ± 0.81 & 4.37 \\
65 & 12:21:51.151 & +04:28:46.942 & 24.28 ± 0.04 & 25.22 ± 0.03 & 39.45 ± 0.59 & 33.06 ± 0.31 & 3.82 \\
66 & 12:21:51.061 & +04:28:24.990 & 24.65 ± 0.07 & 25.78 ± 0.04 & 39.82 ± 1.09 & 12.88 ± 0.62 & 4.71 \\
67 & 12:21:59.769 & +04:28:23.270 & 24.91 ± 0.04 & 26.27 ± 0.04 & 39.83 ± 0.58 & 30.13 ± 0.32 & 1.82 \\
68 & 12:21:55.910 & +04:28:57.816 & 24.64 ± 0.04 & 25.80 ± 0.02 & 40.05 ± 0.51 & 11.95 ± 0.42 & 1.60 \\
69 & 12:21:55.705 & +04:29:28.256 & 24.36 ± 0.03 & 25.44 ± 0.05 & 40.92 ± 1.29 & 38.53 ± 1.19 & 2.26 \\
70 & 12:21:51.805 & +04:27:50.693 & 24.55 ± 0.04 & 25.57 ± 0.03 & 40.99 ± 0.84 & 40.31 ± 0.81 & 1.83 \\
71 & 12:21:57.529 & +04:29:32.191 & 24.85 ± 0.03 & 26.09 ± 0.02 & 41.29 ± 0.51 & 9.69 ± 0.39 & 1.25 \\
72 & 12:21:55.556 & +04:29:19.767 & 24.10 ± 0.04 & 25.19 ± 0.02 & 41.66 ± 0.54 & 21.46 ± 0.31 & 2.12 \\
73 & 12:21:56.817 & +04:27:11.465 & 24.38 ± 0.04 & 25.23 ± 0.03 & 42.56 ± 1.72 & 12.58 ± 1.62 & 7.63 \\
74 & 12:21:52.736 & +04:28:50.161 & 24.59 ± 0.04 & 25.74 ± 0.03 & 42.80 ± 1.28 & 24.40 ± 0.48 & 2.65 \\
75 & 12:21:58.613 & +04:28:07.594 & 24.63 ± 0.03 & 25.93 ± 0.02 & 42.97 ± 1.11 & 7.40 ± 0.41 & 1.50 \\
76 & 12:21:55.808 & +04:27:16.089 & 24.54 ± 0.04 & 25.66 ± 0.02 & 43.27 ± 1.00 & 23.75 ± 0.32 & 2.18 \\
77 & 12:21:57.887 & +04:28:08.721 & 24.43 ± 0.06 & 25.28 ± 0.04 & 43.48 ± 1.13 & 1.57 ± 0.72 & 5.73 \\
78 & 12:21:50.141 & +04:28:53.082 & 24.58 ± 0.03 & 25.78 ± 0.05 & 43.50 ± 1.31 & 38.05 ± 1.44 & 1.35 \\
79 & 12:21:52.484 & +04:27:27.753 & 24.49 ± 0.03 & 25.82 ± 0.03 & 43.49 ± 1.64 & 34.75 ± 0.68 & 2.16 \\
80 & 12:21:54.235 & +04:27:31.145 & 24.40 ± 0.03 & 25.39 ± 0.04 & 43.52 ± 1.83 & 32.93 ± 0.39 & 3.48 \\
81 & 12:21:52.656 & +04:28:47.160 & 24.08 ± 0.03 & 24.85 ± 0.02 & 43.56 ± 1.32 & 33.36 ± 0.40 & 1.88 \\
82 & 12:21:52.489 & +04:27:18.968 & 24.19 ± 0.04 & 25.11 ± 0.02 & 43.57 ± 0.75 & 26.15 ± 0.33 & 3.54 \\
83 & 12:21:50.639 & +04:28:47.148 & 24.44 ± 0.09 & 25.58 ± 0.05 & 43.94 ± 3.11 & 5.83 ± 3.13 & 6.87 \\
84 & 12:21:51.910 & +04:28:43.069 & 24.41 ± 0.04 & 25.37 ± 0.03 & 44.71 ± 0.67 & 28.05 ± 0.34 & 2.53 \\
85 & 12:21:58.158 & +04:28:33.816 & 23.86 ± 0.05 & 24.80 ± 0.02 & 45.18 ± 1.53 & 30.21 ± 0.29 & 2.74 \\
86 & 12:21:56.513 & +04:29:08.928 & 24.41 ± 0.02 & 25.56 ± 0.01 & 46.44 ± 0.34 & 23.25 ± 0.21 & 0.51 \\
87 & 12:21:58.755 & +04:28:16.499 & 24.76 ± 0.06 & 25.77 ± 0.04 & 46.72 ± 1.20 & 16.48 ± 0.73 & 4.20 \\
88 & 12:21:54.718 & +04:29:15.019 & 24.11 ± 0.05 & 25.18 ± 0.03 & 46.86 ± 0.69 & 24.73 ± 0.41 & 1.62 \\
89 & 12:21:58.007 & +04:29:19.719 & 24.56 ± 0.04 & 25.74 ± 0.03 & 47.12 ± 1.34 & 2.97 ± 0.64 & 3.04 \\
90 & 12:21:52.712 & +04:28:35.593 & 24.36 ± 0.05 & 25.24 ± 0.03 & 47.15 ± 1.00 & 14.99 ± 0.57 & 3.39 \\
91 & 12:21:56.523 & +04:29:32.663 & 24.57 ± 0.04 & 26.03 ± 0.04 & 47.19 ± 2.64 & 39.09 ± 1.27 & 0.93 \\
92 & 12:21:57.581 & +04:27:53.402 & 24.17 ± 0.04 & 24.88 ± 0.02 & 47.52 ± 1.65 & 30.81 ± 0.31 & 2.33 \\
93 & 12:21:56.088 & +04:27:47.832 & 24.09 ± 0.03 & 25.16 ± 0.02 & 47.64 ± 0.72 & 0.52 ± 0.57 & 1.10 \\
94 & 12:21:55.744 & +04:28:44.518 & 23.80 ± 0.05 & 24.66 ± 0.03 & 49.20 ± 2.51 & 33.93 ± 0.61 & 4.55 \\
95 & 12:21:50.541 & +04:28:59.003 & 24.11 ± 0.03 & 24.96 ± 0.02 & 49.60 ± 1.59 & 5.33 ± 1.31 & 1.81 \\
96 & 12:21:52.856 & +04:29:27.773 & 24.66 ± 0.04 & 25.76 ± 0.03 & 49.96 ± 1.32 & 19.27 ± 1.29 & 4.75 \\
97 & 12:21:55.531 & +04:29:17.019 & 23.96 ± 0.04 & 25.06 ± 0.02 & 50.63 ± 0.79 & 22.16 ± 0.47 & 2.90 \\
98 & 12:21:51.393 & +04:29:02.296 & 24.44 ± 0.03 & 25.88 ± 0.03 & 50.91 ± 2.28 & 47.21 ± 0.70 & 3.21 \\
99 & 12:21:55.671 & +04:27:05.050 & 24.05 ± 0.04 & 25.04 ± 0.03 & 51.08 ± 1.37 & 13.83 ± 1.25 & 6.62 \\
100 & 12:21:56.226 & +04:27:01.319 & 24.26 ± 0.05 & 25.31 ± 0.03 & 51.11 ± 1.36 & 13.41 ± 0.86 & 6.60 \\
101 & 12:21:55.499 & +04:29:25.395 & 23.98 ± 0.03 & 24.83 ± 0.02 & 51.51 ± 1.31 & 6.43 ± 0.70 & 1.68 \\
102 & 12:21:56.974 & +04:27:27.397 & 24.15 ± 0.04 & 25.22 ± 0.03 & 51.57 ± 1.52 & 14.54 ± 0.84 & 4.41 \\
103 & 12:21:59.145 & +04:27:42.283 & 23.94 ± 0.04 & 24.94 ± 0.03 & 51.67 ± 1.80 & 26.00 ± 0.65 & 4.74 \\
104 & 12:21:53.615 & +04:27:29.220 & 24.01 ± 0.02 & 24.95 ± 0.02 & 52.29 ± 1.44 & 35.50 ± 0.55 & 1.48 \\
105 & 12:21:54.358 & +04:27:03.507 & 24.79 ± 0.04 & 26.19 ± 0.04 & 52.99 ± 3.23 & 45.74 ± 0.74 & 3.18 \\
106 & 12:21:56.117 & +04:29:26.505 & 24.14 ± 0.02 & 25.23 ± 0.02 & 53.38 ± 1.60 & 43.05 ± 0.51 & 0.56 \\
107 & 12:21:52.120 & +04:27:52.187 & 24.05 ± 0.04 & 25.16 ± 0.03 & 54.21 ± 1.78 & 53.03 ± 0.89 & 3.64 \\
108 & 12:21:53.663 & +04:29:30.160 & 24.65 ± 0.04 & 25.92 ± 0.03 & 55.56 ± 3.33 & 46.16 ± 0.68 & 3.08 \\
109 & 12:21:51.271 & +04:28:56.140 & 24.25 ± 0.06 & 25.32 ± 0.05 & 55.73 ± 5.37 & 46.57 ± 1.13 & 10.59 \\
110 & 12:21:50.271 & +04:29:02.666 & 24.49 ± 0.05 & 25.90 ± 0.04 & 56.55 ± 3.68 & 51.43 ± 1.09 & 2.70 \\
111 & 12:21:57.544 & +04:28:41.058 & 23.91 ± 0.03 & 24.98 ± 0.02 & 56.65 ± 2.30 & 39.83 ± 1.11 & 1.10 \\
112 & 12:21:51.132 & +04:29:16.979 & 24.18 ± 0.02 & 25.48 ± 0.02 & 56.98 ± 1.30 & 54.46 ± 0.61 & 1.08 \\
113 & 12:21:56.959 & +04:27:44.300 & 24.10 ± 0.04 & 25.32 ± 0.03 & 59.34 ± 2.35 & 22.43 ± 1.09 & 2.56 \\
114 & 12:21:53.501 & +04:29:33.899 & 23.74 ± 0.04 & 24.99 ± 0.02 & 61.02 ± 2.57 & 4.30 ± 1.82 & 1.56 \\
115 & 12:21:58.834 & +04:27:58.165 & 24.77 ± 0.03 & 26.15 ± 0.03 & 61.55 ± 3.39 & 29.78 ± 0.67 & 1.97 \\
116 & 12:21:56.025 & +04:29:21.437 & 23.89 ± 0.03 & 24.93 ± 0.03 & 61.72 ± 3.95 & 50.63 ± 1.22 & 2.99 \\
117 & 12:21:55.324 & +04:29:27.332 & 23.71 ± 0.03 & 24.53 ± 0.02 & 61.77 ± 2.14 & 50.29 ± 0.71 & 2.63 \\
118 & 12:21:50.569 & +04:28:34.755 & 24.04 ± 0.04 & 25.49 ± 0.03 & 63.35 ± 2.36 & 0.81 ± 1.76 & 3.43 \\
119 & 12:21:53.061 & +04:29:20.687 & 23.68 ± 0.03 & 24.59 ± 0.03 & 63.43 ± 2.83 & 34.48 ± 0.91 & 5.09 \\
120 & 12:21:58.081 & +04:28:26.644 & 23.83 ± 0.03 & 24.69 ± 0.02 & 64.00 ± 1.65 & 22.85 ± 1.57 & 1.17 \\
121 & 12:21:55.285 & +04:27:06.767 & 23.47 ± 0.05 & 24.55 ± 0.01 & 65.92 ± 1.94 & 35.64 ± 0.87 & 1.52 \\
122 & 12:21:49.717 & +04:28:45.428 & 23.68 ± 0.04 & 24.83 ± 0.04 & 66.05 ± 8.39 & 46.29 ± 1.84 & 8.02 \\
123 & 12:21:54.567 & +04:29:29.406 & 23.56 ± 0.03 & 24.54 ± 0.03 & 66.50 ± 4.13 & 50.10 ± 1.31 & 5.23 \\
124 & 12:21:55.068 & +04:29:12.310 & 23.99 ± 0.03 & 25.03 ± 0.02 & 68.17 ± 3.81 & 53.78 ± 1.10 & 1.83 \\
125 & 12:21:51.227 & +04:27:25.572 & 23.72 ± 0.02 & 25.17 ± 0.02 & 73.32 ± 4.28 & 62.94 ± 1.19 & 1.21 \\
126 & 12:21:56.383 & +04:29:20.021 & 24.01 ± 0.04 & 25.24 ± 0.04 & 75.69 ± 7.39 & 27.88 ± 1.59 & 3.35 \\
127 & 12:21:56.170 & +04:29:29.822 & 23.24 ± 0.04 & 24.00 ± 0.02 & 77.92 ± 7.27 & 38.20 ± 2.08 & 3.12 \\
128 & 12:21:58.194 & +04:28:33.806 & 22.90 ± 0.03 & 23.96 ± 0.03 & 86.31 ± 8.39 & 75.34 ± 3.27 & 3.42 \\
129 & 12:21:57.199 & +04:27:53.671 & 22.93 ± 0.03 & 23.78 ± 0.02 & 86.48 ± 6.66 & 29.58 ± 1.32 & 2.59 \\
130 & 12:21:56.431 & +04:28:49.211 & 23.68 ± 0.01 & 24.87 ± 0.02 & 93.27 ± 4.33 & 19.15 ± 0.97 & 0.45 \\
\enddata
    \tablecomments{The ID's of each Cepheid candidate correspond to the ID's of each light curve in Figure \ref{fig:n4303_lcs}. The phases here are given in days with respect to the first visit. $\chi^2 / \nu$ is the reduced $\chi^2$ value for the F555W Cepheid template fit.}
\end{deluxetable*}

\clearpage

\begin{figure*}[!ht]
    \centering
    \includegraphics[width=\columnwidth]{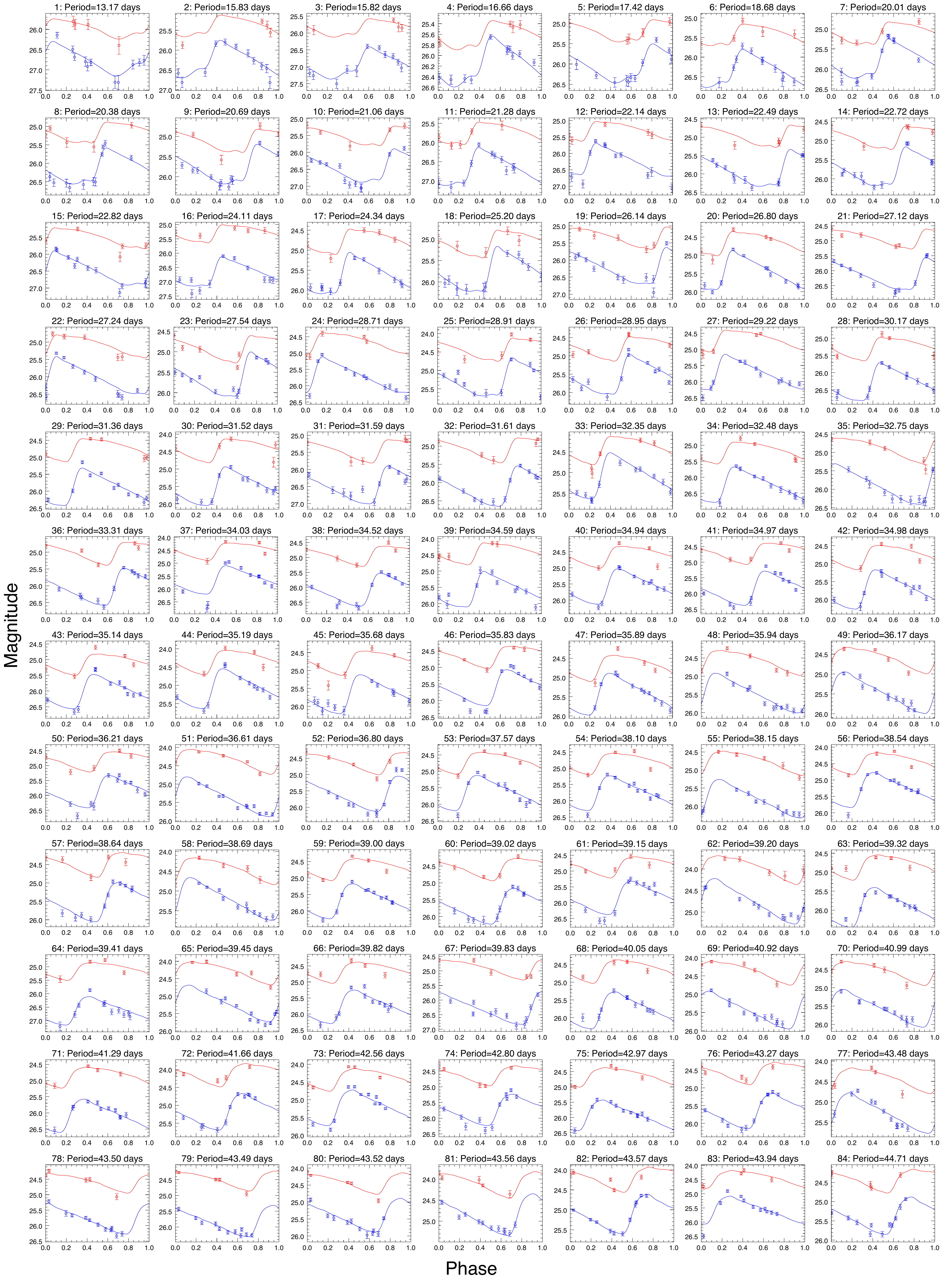}
\end{figure*}

\clearpage

\section{NGC\,1068} \label{sec:app_b}

\begin{figure*}[!ht]
    \centering
    \includegraphics[width=\columnwidth]{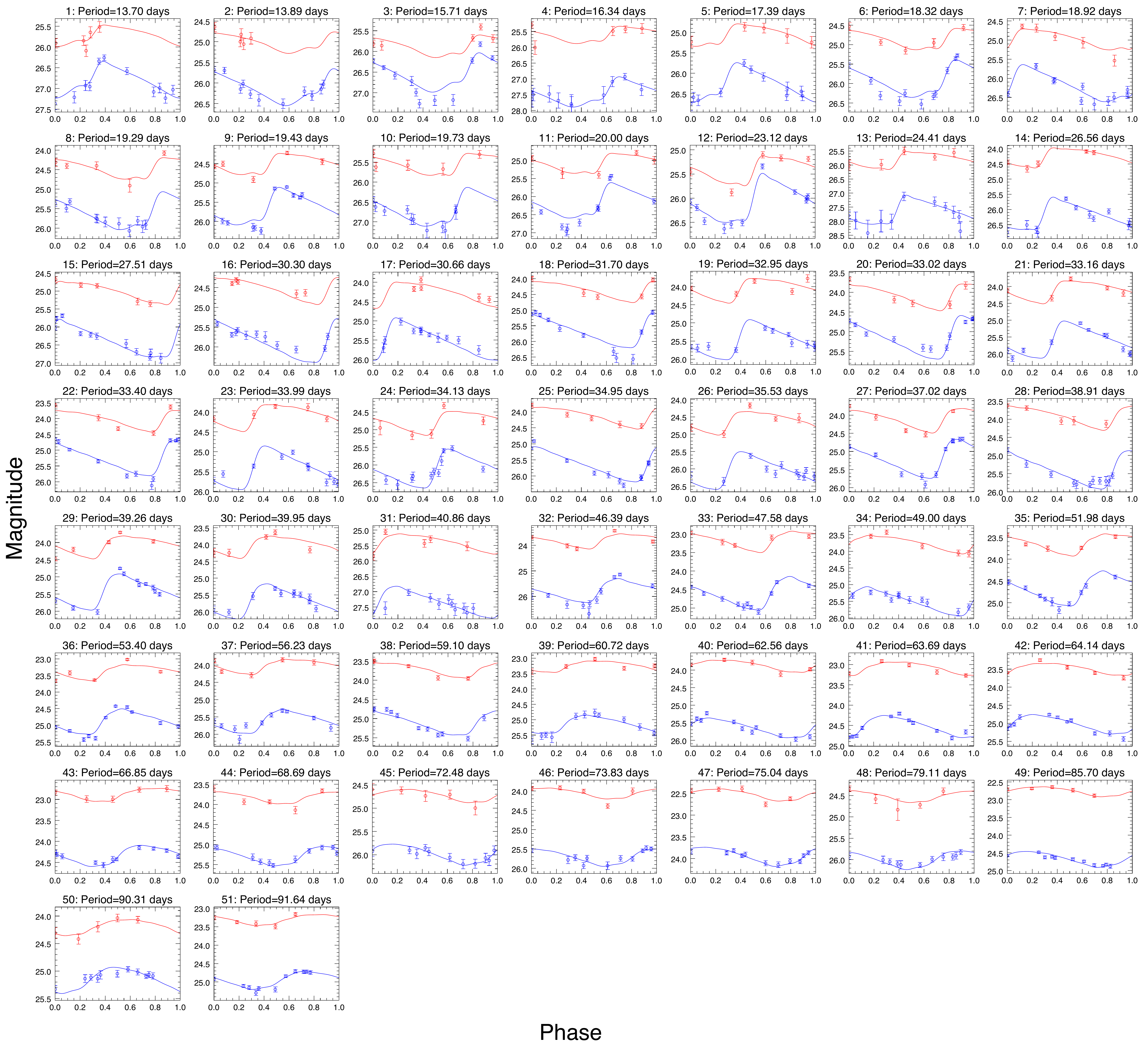}
    \caption{Light curves for 51 final Cepheid candidates in NGC\,1068. The data points show the measurements in F555W (blue) and F814W (red), while the solid lines represent the best Cepheid template fits to the data.}
    \label{fig:n1068_lcs}
\end{figure*}

As in Section \ref{sec:app_a}, the Cepheid candidates in Table \ref{tab:n1068_ceph} and Figure \ref{fig:n1068_lcs} represent our final $3\sigma$-clipped sample for NGC\,1068. Sources with periods below the point of completeness ($P>25$ days, as determined by Figure \ref{fig:comp}) are also included.

\startlongtable
\begin{deluxetable*}{rccccccc}
    \tablecaption{Best-fit Light Curve Parameters for Final Cepheid Candidates in NGC\,1068}

\label{tab:n1068_ceph}

\tablehead{
\colhead{ID} &
\colhead{RA} &
\colhead{Dec} &
\colhead{$m_{\rm F555W}$} &
\colhead{$m_{\rm F814W}$} &
\colhead{$P$} &
\colhead{Phase} &
\colhead{$\chi^2 / \nu$} \\ 
\colhead{} &
\colhead{(hh:mm:ss)} &
\colhead{(dd:mm:ss)} &
\colhead{(mag)} &
\colhead{(mag)} &
\colhead{(days)} &
\colhead{(days)} &
\colhead{}
}

\startdata
1 & 02:42:43.828 & $-$00:01:27.473 & 25.72 ± 0.07 & 26.83 ± 0.04 & 13.70 ± 0.27 & 3.13 ± 0.94 & 1.48 \\
2 & 02:42:41.336 & $-$00:01:26.882 & 25.02 ± 0.06 & 26.15 ± 0.03 & 13.89 ± 0.11 & 11.13 ± 0.35 & 1.26 \\
3 & 02:42:46.415 & $-$00:01:18.621 & 25.88 ± 0.09 & 26.57 ± 0.07 & 15.71 ± 0.39 & 11.02 ± 1.07 & 5.22 \\
4 & 02:42:40.588 & $-$00:01:36.045 & 25.61 ± 0.07 & 27.47 ± 0.06 & 16.34 ± 0.39 & 9.14 ± 1.03 & 1.04 \\
5 & 02:42:41.153 & $-$00:01:22.600 & 25.07 ± 0.04 & 26.28 ± 0.03 & 17.39 ± 0.24 & 4.06 ± 0.45 & 0.55 \\
6 & 02:42:41.220 & $-$00:00:01.335 & 24.86 ± 0.07 & 25.93 ± 0.04 & 18.32 ± 0.50 & 13.31 ± 0.58 & 3.02 \\
7 & 02:42:42.910 & $-$00:00:24.682 & 24.98 ± 0.06 & 26.19 ± 0.05 & 18.92 ± 0.43 & 14.59 ± 0.79 & 2.25 \\
8 & 02:42:38.743 & $-$00:01:10.896 & 24.47 ± 0.05 & 25.62 ± 0.05 & 19.29 ± 0.43 & 14.25 ± 0.84 & 1.50 \\
9 & 02:42:38.895 & $-$00:00:03.623 & 24.51 ± 0.06 & 25.67 ± 0.04 & 19.43 ± 0.35 & 7.77 ± 0.93 & 4.41 \\
10 & 02:42:41.015 & $-$00:01:48.774 & 25.55 ± 0.06 & 26.69 ± 0.04 & 19.73 ± 0.34 & 12.57 ± 0.43 & 1.34 \\
11 & 02:42:44.100 & $-$00:01:23.543 & 25.08 ± 0.10 & 26.17 ± 0.05 & 20.00 ± 0.28 & 10.42 ± 0.53 & 6.73 \\
12 & 02:42:45.499 & $-$00:00:15.180 & 25.41 ± 0.05 & 26.08 ± 0.05 & 23.12 ± 0.44 & 11.02 ± 1.03 & 2.70 \\
13 & 02:42:43.513 & $-$00:01:32.601 & 25.81 ± 0.06 & 27.68 ± 0.05 & 24.41 ± 0.73 & 8.83 ± 1.19 & 0.74 \\
14 & 02:42:41.467 & $-$00:01:34.264 & 24.26 ± 0.06 & 26.20 ± 0.05 & 26.56 ± 0.80 & 7.22 ± 1.35 & 2.10 \\
15 & 02:42:44.531 & $-$00:00:38.679 & 25.02 ± 0.04 & 26.40 ± 0.03 & 27.51 ± 0.64 & 22.38 ± 0.51 & 1.10 \\
16 & 02:42:43.356 & $-$00:00:22.337 & 24.53 ± 0.05 & 25.89 ± 0.04 & 30.30 ± 0.48 & 21.65 ± 0.34 & 3.29 \\
17 & 02:42:41.198 & $-$00:00:01.252 & 24.28 ± 0.05 & 25.53 ± 0.04 & 30.66 ± 0.56 & 21.67 ± 0.46 & 2.33 \\
18 & 02:42:42.961 & $-$00:00:39.990 & 24.36 ± 0.06 & 25.69 ± 0.04 & 31.70 ± 0.93 & 27.84 ± 0.87 & 2.68 \\
19 & 02:42:39.105 & $-$00:01:09.252 & 24.05 ± 0.05 & 25.50 ± 0.04 & 32.95 ± 1.18 & 12.39 ± 1.11 & 2.43 \\
20 & 02:42:39.463 & $-$00:01:05.191 & 24.06 ± 0.07 & 25.18 ± 0.03 & 33.02 ± 0.52 & 26.37 ± 0.39 & 3.50 \\
21 & 02:42:40.112 & $-$00:00:11.071 & 24.11 ± 0.06 & 25.61 ± 0.04 & 33.16 ± 1.19 & 11.82 ± 1.16 & 4.03 \\
22 & 02:42:41.460 & $-$00:01:24.792 & 24.02 ± 0.07 & 25.29 ± 0.04 & 33.40 ± 1.15 & 28.13 ± 0.91 & 4.28 \\
23 & 02:42:43.068 & $-$00:00:39.695 & 24.09 ± 0.06 & 25.44 ± 0.03 & 33.99 ± 1.10 & 10.52 ± 1.08 & 3.00 \\
24 & 02:42:43.172 & $-$00:00:42.286 & 24.76 ± 0.10 & 26.13 ± 0.07 & 34.13 ± 0.56 & 31.89 ± 0.47 & 3.85 \\
25 & 02:42:40.611 & $-$00:01:44.783 & 24.13 ± 0.06 & 25.66 ± 0.04 & 34.95 ± 0.46 & 32.29 ± 0.31 & 3.88 \\
26 & 02:42:38.892 & $-$00:01:25.492 & 24.62 ± 0.06 & 26.07 ± 0.05 & 35.53 ± 1.56 & 10.53 ± 0.66 & 2.62 \\
27 & 02:42:43.261 & $-$00:00:28.701 & 24.09 ± 0.04 & 25.22 ± 0.03 & 37.02 ± 0.70 & 27.51 ± 0.63 & 3.37 \\
28 & 02:42:38.670 & $-$00:01:03.271 & 23.91 ± 0.05 & 25.40 ± 0.04 & 38.91 ± 1.02 & 33.07 ± 0.39 & 1.97 \\
29 & 02:42:38.017 & $-$00:00:04.772 & 24.08 ± 0.06 & 25.45 ± 0.05 & 39.26 ± 1.62 & 15.09 ± 1.10 & 7.77 \\
30 & 02:42:38.762 & $-$00:00:40.367 & 24.02 ± 0.06 & 25.70 ± 0.04 & 39.95 ± 1.58 & 11.79 ± 1.65 & 2.03 \\
31 & 02:42:36.887 & $-$00:00:38.555 & 25.39 ± 0.07 & 27.35 ± 0.07 & 40.86 ± 1.52 & 2.11 ± 1.41 & 1.76 \\
32 & 02:42:40.861 & $-$00:01:11.386 & 23.79 ± 0.06 & 25.77 ± 0.06 & 46.39 ± 1.93 & 25.45 ± 1.17 & 4.12 \\
33 & 02:42:41.435 & $-$00:01:27.152 & 23.16 ± 0.03 & 24.61 ± 0.02 & 47.58 ± 1.42 & 30.73 ± 0.29 & 1.35 \\
34 & 02:42:41.112 & $-$00:01:06.934 & 23.75 ± 0.05 & 25.50 ± 0.04 & 49.00 ± 1.19 & 1.95 ± 0.75 & 2.00 \\
35 & 02:42:43.640 & $-$00:00:26.371 & 23.63 ± 0.02 & 24.68 ± 0.02 & 51.98 ± 1.31 & 32.00 ± 0.27 & 1.21 \\
36 & 02:42:38.483 & $-$00:01:07.884 & 23.38 ± 0.06 & 24.90 ± 0.05 & 53.40 ± 2.80 & 20.46 ± 1.24 & 12.05 \\
37 & 02:42:41.379 & $-$00:00:09.759 & 24.02 ± 0.05 & 25.64 ± 0.03 & 56.23 ± 2.46 & 22.46 ± 0.84 & 2.58 \\
38 & 02:42:43.481 & $-$00:00:22.368 & 23.71 ± 0.04 & 25.12 ± 0.03 & 59.10 ± 3.67 & 49.50 ± 1.86 & 3.21 \\
39 & 02:42:41.059 & $-$00:01:23.397 & 23.25 ± 0.04 & 25.15 ± 0.04 & 60.72 ± 4.38 & 16.88 ± 1.79 & 1.93 \\
40 & 02:42:39.462 & $-$00:01:38.151 & 23.86 ± 0.04 & 25.64 ± 0.03 & 62.56 ± 1.74 & 61.65 ± 1.09 & 2.43 \\
41 & 02:42:41.061 & $-$00:00:01.724 & 23.08 ± 0.03 & 24.53 ± 0.02 & 63.69 ± 2.38 & 7.99 ± 0.97 & 2.86 \\
42 & 02:42:43.648 & $-$00:00:50.941 & 23.48 ± 0.03 & 25.02 ± 0.02 & 64.14 ± 1.14 & 2.91 ± 0.85 & 2.61 \\
43 & 02:42:38.955 & $-$00:00:44.256 & 22.87 ± 0.03 & 24.33 ± 0.02 & 66.85 ± 2.34 & 34.72 ± 0.66 & 1.21 \\
44 & 02:42:43.042 & $-$00:00:37.908 & 23.82 ± 0.04 & 25.28 ± 0.03 & 68.69 ± 5.16 & 49.04 ± 1.67 & 1.61 \\
45 & 02:42:40.974 & $-$00:00:09.255 & 24.72 ± 0.04 & 25.98 ± 0.02 & 72.48 ± 3.24 & 70.62 ± 1.63 & 0.68 \\
46 & 02:42:41.853 & $-$00:01:32.254 & 24.05 ± 0.04 & 25.70 ± 0.04 & 73.83 ± 8.31 & 60.02 ± 2.03 & 1.95 \\
47 & 02:42:39.674 & $-$00:01:04.311 & 22.52 ± 0.03 & 23.94 ± 0.02 & 75.04 ± 4.07 & 68.65 ± 1.18 & 1.44 \\
48 & 02:42:41.203 & $-$00:01:32.948 & 24.51 ± 0.05 & 26.00 ± 0.04 & 79.11 ± 13.74 & 55.39 ± 2.90 & 1.47 \\
49 & 02:42:38.652 & $-$00:01:28.973 & 22.76 ± 0.03 & 24.67 ± 0.02 & 85.70 ± 4.06 & 82.38 ± 3.01 & 1.95 \\
50 & 02:42:40.186 & $-$00:01:26.054 & 24.19 ± 0.04 & 25.15 ± 0.02 & 90.31 ± 5.67 & 26.20 ± 1.34 & 1.01 \\
51 & 02:42:42.400 & $-$00:01:06.240 & 23.30 ± 0.03 & 24.94 ± 0.02 & 91.64 ± 3.98 & 49.90 ± 1.29 & 1.21 \\
\enddata

    \tablecomments{The ID's of each Cepheid candidate correspond to the ID's of each light curve in Figure \ref{fig:n1068_lcs}. The phases here are given in days with respect to the first visit. $\chi^2 / \nu$ is the reduced $\chi^2$ value for the F555W Cepheid template fit.}
\end{deluxetable*}

\clearpage

\bibliography{cepheids_ref}{}
\bibliographystyle{aasjournal}

\end{document}